\documentclass{elsart}

\usepackage{amssymb}
\usepackage{graphicx}

\begin{document}

\begin{frontmatter}

\title{Seniority in quantum many-body systems.\\
I. Identical particles in a single shell}


\author[Caen]{P.~Van Isacker},
\ead{isacker@ganil.fr}
\author[Koeln]{S.~Heinze}

\address[Caen]{Grand Acc\'el\'erateur National d'Ions Lourds,
CEA/DSM--CNRS/IN2P3, BP~55027, F-14076 Caen Cedex 5, France}
\address[Koeln]{Institut f\"ur Kernphysik der Universit\"at zu K\"oln,
50937 K\"oln, Germany}

\begin{abstract}
A discussion of the seniority quantum number in many-body systems is presented.
The analysis is carried out for bosons and fermions simultaneously
but is restricted to identical particles occupying a single shell.
The emphasis of the paper is
on the possibility of {\em partial} conservation of seniority
which turns out to be a peculiar property of spin-9/2 fermions
but prevalent in systems of interacting bosons of any spin.
Partial conservation of seniority is
at the basis of the existence of seniority isomers,
frequently observed in semi-magic nuclei,
and also gives rise to peculiar selection rules in one-nucleon transfer reactions.
\end{abstract}

\end{frontmatter}

\section{Introduction}
\label{s_int}
The seniority quantum number was introduced by Racah
for the classification of electrons in an $\ell^N$ configuration
where it appears as a label
additional to the total orbital angular momentum $L$,
the total spin $S$
and the total angular momentum $J$~\cite{Racah43}.
About ten years after its introduction by Racah
it was adopted in nuclear physics
for the $jj$-coupling classification of nucleons
in a single-$j$ shell~\cite{Racah52,Flowers52}.
Seniority refers to the number of particles that are not in pairs
coupled to angular momentum $J=0$.
The seniority quantum number is usually denoted by $v$,
from the Hebrew word for seniority, `vet(h)ek'~\cite{Shalit63,Talmi93}.
In nuclear physics the concept has proven extremely useful,
especially in semi-magic nuclei
where only one type of nucleon (neutron or proton) is active
and where seniority turns out to be conserved
to a good approximation.

Seniority can be given a group-theoretical definition
starting from the Lie algebra U($2j+1$)
which contains all (infinitesimal) unitary transformations
among the $2j+1$ single-particle states
$|jm_j\rangle$ with $m_j=-j,-j+1,\dots,+j$,
where $j$ is the angular momentum (henceforth referred to as spin)
carried by the particle
which is integer for bosons and half-odd-integer for fermions.
A system of $N$ identical particles
({\it i.e.}, no other internal degrees of freedom for the particles besides their spin)
is characterized by the symmetric representation $[N]$ of U($2j+1$)
in the case of bosons
or the anti-symmetric representation $[1^N]\equiv[1,1,\dots,1]$
in the case of fermions.
Seniority arises as a label $v$ associated with a subalgebra of U($2j+1$),
either the orthogonal algebra SO($2j+1$), if $2j+1$ is odd (bosons),
or the (unitary) symplectic algebra Sp($2j+1$), if $2j+1$ is even (fermions).
Both SO($2j+1$) and Sp($2j+1$) contain the rotation algebra as a subalgebra
which shall be denoted as SO(3) and SU(2), respectively,
to indicate that the total angular momentum $J$ must by integer for bosons
whereas it can be integer or half-odd-integer for fermions,
depending on $N$ being even or odd.
Finally, any many-particle state is characterized
by the projection $M_J$ of the total angular momentum $J$ associated with SO(2).

The seniority classification can be summarized as
\begin{equation}
\begin{array}{ccccccccc}
{\rm U}(2j+1)&\supset&{\rm SO}(2j+1)&\supset&\cdots &\supset&{\rm SO}(3)&\supset&{\rm SO}(2)\\
\downarrow&&\downarrow&&\downarrow&&\downarrow&&\downarrow\\[0mm]
[N]&&v&&\alpha&&J&&M_J
\end{array},
\label{e_clasb}
\end{equation}
and
\begin{equation}
\begin{array}{ccccccccc}
{\rm U}(2j+1)&\supset&{\rm Sp}(2j+1)&\supset&\cdots&\supset&{\rm SU}(2)&\supset&{\rm SO}(2)\\
\downarrow&&\downarrow&&\downarrow&&\downarrow&&\downarrow\\[0mm]
[1^N]&&v&&\alpha&&J&&M_J
\end{array},
\label{e_clasf}
\end{equation}
for bosons and fermions, respectively.
In general, this classification is not complete,
as indicated by the dots in the above equations.
The allowed values of $v$ are $v=N,N-2,\dots,1$ or 0,
as can be obtained from the ${\rm U}(2j+1)\supset{\rm SO}(2j+1)$
or ${\rm U}(2j+1)\supset{\rm Sp}(2j+1)$
branching rules~\cite{Wybourne74}.
The allowed values of the total spin $J$
are obtained from the ${\rm SO}(2j+1)\supset{\rm SO}(3)$
or ${\rm Sp}(2j+1)\supset{\rm SU}(2)$  branching rules,
which in general require a multiplicity label $\alpha$.
Alternatively, seniority can be introduced
via the quasi-spin formalism~\cite{Kerman61,Helmers61}
where it arises as a label associated with the Lie algebras SU(1,1) or SU(2)
for bosons or fermions, respectively.
This alternative definition is particularly valuable
for generalizations towards different shells or several types of particles
but it is not needed here.

This paper is concerned with the seniority classification
of a system of $N$ identical particles with spin $j$ 
interacting through a general rotationally invariant two-body force.
Symmetry arguments then dictate that the eigenstates of the hamiltonian
carry good angular momentum $J$, as in the classifications~(\ref{e_clasb}) and~(\ref{e_clasf}).
The central question addressed in this paper is
what conditions are required for seniority $v$ to be a good quantum number
for all or for part of the eigenstates.

To arrive at a more precise formulation of this question,
let us introduce the following notation.
A rotationally invariant two-body interaction $\hat V$ between the particles
is specified by its $\lfloor j+1\rfloor$ matrix elements
$\nu_\lambda\equiv\langle j^2;\lambda m_\lambda
|\hat V|j^2;\lambda m_\lambda\rangle$
(where $\lfloor x\rfloor$ is the largest integer
smaller than or equal to $x$).
The notation $|j^2;\lambda m_\lambda\rangle$
implies a normalized two-particle state
with total angular momentum $\lambda$ and projection $m_\lambda$
which can take the values $\lambda=0,2,\dots,2p$,
$m_\lambda=-\lambda,-\lambda+1,\dots,+\lambda$,
where $2p=2j$ for bosons and $2p=2j-1$ for fermions.
Since the interaction is rotationally invariant,
there is no dependence on the label $m_\lambda$
which shall be suppressed henceforth.
The interaction can then be written as
$\hat V=\sum_\lambda \nu_\lambda\hat V_\lambda$
where $\hat V_\lambda$ is the operator defined through
$\langle j^2;\lambda'|\hat V_\lambda|j^2;\lambda''\rangle=
\delta_{\lambda\lambda'}\delta_{\lambda\lambda''}$.

A precise formulation can now be given
of the question that will be addressed in this paper:
What conditions should the matrix elements $\nu_\lambda$ satisfy
for the interaction $\hat V$ to conserve seniority,
either completely or partially?
It is important to appreciate that these conditions are weaker
than those required for complete solvability
on the basis of a dynamical symmetry.
To make this point clear,
sufficient conditions of solvability associated with a dynamical symmetry
are derived in Sect.~\ref{s_solv}.
The conditions for {\em complete} seniority conservation
are known since long.
For completeness, a brief reminder of them is given in Sect.~\ref{s_sen3}
by analyzing the three-particle case.
A surprising consequence of these conditions
is that they lead to diophantine equations in the spin $j$ of the particles
and the component $\lambda$ of the interaction.
With the same procedure as in Sect.~\ref{s_sen3},
the four-particle case is analyzed in Sect.~\ref{s_sen4},
revealing the existence 
of a {\em partial} seniority conservation in the $j=9/2$ shell.
Although this is found to be an exceptional situation for fermions,
it occurs frequently for boson systems
as is shown in Sect.~\ref{s_senbos}.
Applications of the seniority formalism in fermionic systems
are presented in Sect.~\ref{s_appfer}.
Finally, in Sect.~\ref{s_conc}
the conclusions of this work are formulated.

\section{Solvable interactions}
\label{s_solv}
A class of solvable interactions
can be found by requiring the existence of a dynamical symmetry
which can be viewed
as a generalization and refinement
of the concept of symmetry~\cite{Frank94,Iachello06}.
A dynamical symmetry occurs if the hamiltonian is written
in terms of Casimir operators of a set of nested algebras.
Its hallmarks are
(i) solvability of the complete spectrum,
(ii) existence of exact quantum numbers for all eigenstates
and (iii) pre-determined structure  of the eigenfunctions,
independent of the parameters in the hamiltonian.

A general one- plus two-body hamiltonian
for a system of identical interacting particles considered here
is given by
\begin{equation}
\hat H=\epsilon\hat N+\sum_\lambda\nu_\lambda\hat V_\lambda,
\label{e_ham}
\end{equation}
where $\epsilon$ is the single-particle energy.
If the hamiltonian can be written as a linear combination
of the Casimir operators of the algebras
appearing in Eqs.~(\ref{e_clasb}) and~(\ref{e_clasf}),
then the labels $N$, $v$ and $J$
are good quantum numbers for all eigenstates.
In this case the hamiltonian has the form
\begin{equation}
\hat H_{\rm ds}^{\rm b}=
x_1\hat C_1[{\rm U}(n)]+
x_2\hat C_2[{\rm U}(n)]+
x_3\hat C_2[{\rm SO}(n)]+
x_4\hat C_2[{\rm SO}(3)],
\label{e_hamb}
\end{equation}
or
\begin{equation}
\hat H_{\rm ds}^{\rm f}=
x_1\hat C_1[{\rm U}(n)]+
x_2\hat C_2[{\rm U}(n)]+
x_3\hat C_2[{\rm Sp}(n)]+
x_4\hat C_2[{\rm SU}(2)],
\label{e_hamf}
\end{equation}
for bosons or fermions, respectively,
where the notation $n\equiv2j+1$ is used
and $\hat C_i[G]$ denotes the Casimir operator of order $i$ of the algebra $G$.
By writing the Casimir operator in terms of $\hat N$ and $\hat V_\lambda$,
one obtains a (possibly overcomplete) system of linear equations
in the coefficients $\epsilon$ and $\nu_\lambda$ of the general hamiltonian.
From a simple counting argument
it is clear that these equations admit a solution for $j=0$, 1/2, 1, 3/2, 2 and 5/2.
For $j>5/2$ there are more coefficients than there are Casimir operators.
The system of equations then becomes overcomplete,
leading to conditions on the coefficients $\nu_\lambda$.
There will be one condition for $j=3$ or 7/2,
two conditions for $j=4$ or 9/2, and so on.
The system of equations can be written in general as
\begin{eqnarray}
x_1&=&\epsilon-
\frac{1}{2n}\nu_0+
\frac{n^4+n^3-41n^2-n+40}{56n}\nu_2-
\frac{n^4+n^3-13n^2-n+12}{56n}\nu_4,
\nonumber\\
x_2&=&\frac{1}{2n}\left(\nu_0-
\frac{n^3-41n+40}{28}\nu_2+
\frac{n^3-13n+12}{28}\nu_4\right),
\nonumber\\
x_3&=&\frac{1}{2n}\left(-\nu_0+
\frac{10}{7}\nu_2-\frac{3}{7}\nu_4\right),
\nonumber\\
x_4&=&\frac{\nu_4-\nu_2}{14}=\frac{\nu_\lambda-\nu_2}{\lambda(\lambda+1)-6},
\end{eqnarray}
and
\begin{eqnarray}
x_1&=&\epsilon+\frac{1}{2n}\nu_0-
\frac{n^4-n^3-41n^2+n+40}{56n}\nu_2+
\frac{n^4-n^3-13n^2+n+12}{56n}\nu_4,
\nonumber\\
x_2&=&\frac{1}{2n}\left(\nu_0+
\frac{n^3-41n-40}{28}\nu_2-
\frac{n^3-13n-12}{28}\nu_4\right),
\nonumber\\
x_3&=&\frac{1}{2n}\left(-\nu_0+
\frac{10}{7}\nu_2-\frac{3}{7}\nu_4 \right),
\nonumber\\
x_4&=&\frac{\nu_4-\nu_2}{14}=
\frac{\nu_\lambda-\nu_2}{\lambda(\lambda+1)-6},
\end{eqnarray}
for bosons and fermions, respectively.
In each case the conditions on the interactions $\nu_\lambda$
follow from the last equation 
which is the same for bosons and fermions,
\begin{equation}
\nu_\lambda=
\frac{20-\lambda(\lambda+1)}{14}\nu_2-
\frac{6-\lambda(\lambda+1)}{14}\nu_4,
\qquad
\lambda=6,8,10,\dots
\label{e_solv}
\end{equation}
These are sufficient conditions on the $\nu_\lambda$
for the hamiltonian to have a dynamical symmetry,
resulting in complete solvability of the spectrum.
In the following section the weaker conditions are reviewed
which are needed for complete conservation of seniority.

\section{Seniority conservation for three identical particles}
\label{s_sen3}
The conditions for complete seniority conservation
are known since long for fermions (see, {\it e.g.}, Refs.~\cite{Shalit63,Talmi93})
and can be derived from the analysis of a system of three particles.
This section presents a succinct derivation
of the conditions for seniority conservation for bosons as well as fermions,
to prepare the ground for the analysis of a four-particle system,
presented in Sect.~\ref{s_sen4}.

\subsection{Conditions for seniority conservation}
\label{ss_con3}
Let us recall a few elementary properties
of (anti-)symmetric three-particle states~\cite{Shalit63,Talmi93}.
A three-particle state can be written as $|j^2(R)j;J\rangle$
where two particles are first coupled to angular momentum $R$
which is subsequently coupled to total angular momentum $J$.
This state is not (anti-)symmetric in all three particles;
it can be made so by applying the (anti-)symmetry operator $\hat P$,
\begin{equation}
|j^3[I]J\rangle
\propto{\hat P}|j^2(I)j;J\rangle
=\sum_R\;
[j^2(R)j;J|\}j^3[I]J]\;
|j^2(R)j;J\rangle,
\label{e_3pstate}
\end{equation}
where $[j^2(R)j;J|\}j^3[I]J]$ is a three-to-two-particle
coefficient of fractional par\-entage (CFP).
The notation in round brackets in $|j^2(R)j;J\rangle$
implies coupling of two particles
to intermediate angular momentum $R$.
On the other hand, the square brackets $[I]$
label a three-particle state
and indicate that it has been obtained
after (anti-)symmetrization of $|j^2(I)j;J\rangle$.
The label $[I]$ defines an overcomplete, non-orthogonal basis,
that is, not all $|j^3[I]J\rangle$ states with $I=0,2,\dots,2p$
are independent.

The three-to-two-particle  CFP is known in closed form,
\begin{equation}
[j^2(R)j;J|\}j^3[I]J]=
\frac{1}{\sqrt{{N}_{jJ}^I}}
\left(\delta_{RI}+2\sqrt{(2R+1)(2I+1)}
\Biggl\{\begin{array}{ccc}j&j&R\\ J&j&I\end{array}\Biggr\}\right),
\label{e_32cfp}
\end{equation}
with the normalization coefficient
\begin{equation}
{N}_{jJ}^I=
3\left(1+2(2I+1)
\Biggl\{\begin{array}{ccc}j&j&I\\ J&j&I\end{array}\Biggr\}\right),
\label{e_32norm}
\end{equation}
where the symbol between curly brackets
is a Racah coefficient~\cite{Shalit63,Talmi93}.
Both the overlap matrix
and the matrix element of the operator $\hat V_\lambda$
can be expressed in terms of the CFPs,
\begin{eqnarray}
\langle j^3[I]J|j^3[L]J\rangle&=&
\sum_R\;
[j^2(R)j;J|\}j^3[I]J]\;
[j^2(R)j;J|\}j^3[L]J],
\label{e_overlap3s}\\
\langle j^3[I]J|\hat V_\lambda|j^3[L]J\rangle&=&
3\;[j^2(\lambda)j;J|\}j^3[I]J]\;
[j^2(\lambda)j;J|\}j^3[L]J].
\label{e_hmat3}
\end{eqnarray}
With use of properties of the Racah coefficient
the sum over the CFPs
in the expression for the overlap matrix
can be carried out,
\begin{eqnarray}
\langle j^3[I]J|j^3[L]J\rangle&=&
\frac{3}{\sqrt{{N}_{jJ}^I{N}_{jJ}^L}}
\left(\delta_{IL}+2\sqrt{(2I+1)(2L+1)}
\Biggl\{\begin{array}{ccc}j&j&I\\ J&j&L\end{array}\Biggr\}\right)
\nonumber\\
&=&
\frac{3}{\sqrt{{N}_{jJ}^I}}
[j^2(I)j;J|\}j^3[L]J],
\label{e_overlap3}
\end{eqnarray}
leading to closed expressions for both the overlap matrix
and the matrix element of the operator $\hat V_\lambda$.

\subsection{Diophantine equations for seniority conservation}
\label{ss_dio}
Let us begin with the following simpler problem.
Can one find the condition
for a single component $\hat V_\lambda$ of the interaction
to conserve seniority for a given particle angular momentum $j$?
Let us first establish a necessary condition for seniority conservation~\cite{Shalit63,Talmi93}.
By definition the seniority $v=1$ three-particle state is
\begin{equation}
|j^3,v=1,J\rangle=|j^3[0]J\rangle,
\label{e_v1}
\end{equation}
where the total angular momentum $J$
must be equal to the angular momentum $j$ of the individual particles. 
A seniority $v=3$ state originates from a different parent state
({\it i.e.}, it has $I\neq0$)
and is defined to be orthogonal to the seniority $v=1$ state.
Hence
\begin{equation}
|j^3[I],v=3,J\rangle=
|j^3[I]J\rangle-
\langle j^3[0]J|j^3[I]J\rangle|j^3[0]J\rangle,
\quad I\neq0.
\label{e_v3}
\end{equation}
Seniority conservation for $\hat V_\lambda$ requires
$\langle j^3,v=1,J|\hat V_\lambda|j^3[I],v=3,J\rangle=0$
or
\begin{equation}
\frac{\langle j^3[0]J|\hat V_\lambda|j^3[I]J\rangle}
{\langle j^3[0]J|\hat V_\lambda|j^3[0]J\rangle}=
\langle j^3[0]J|j^3[I]J\rangle.
\label{e_con3}
\end{equation}
With use of the expressions~(\ref{e_hmat3}) and~(\ref{e_overlap3})
this condition reduces to
\begin{equation}
\frac{[j^2(\lambda)j;J|\}j^3[I]J]}
{[j^2(\lambda)j;J|\}j^3[0]J]}=
\frac{3}{\sqrt{{N}_{jj}^0}}
[j^2(0)j;J|\}j^3[I]J].
\label{e_con3a}
\end{equation}

\begin{figure*}
\begin{center}
\includegraphics[width=9cm]{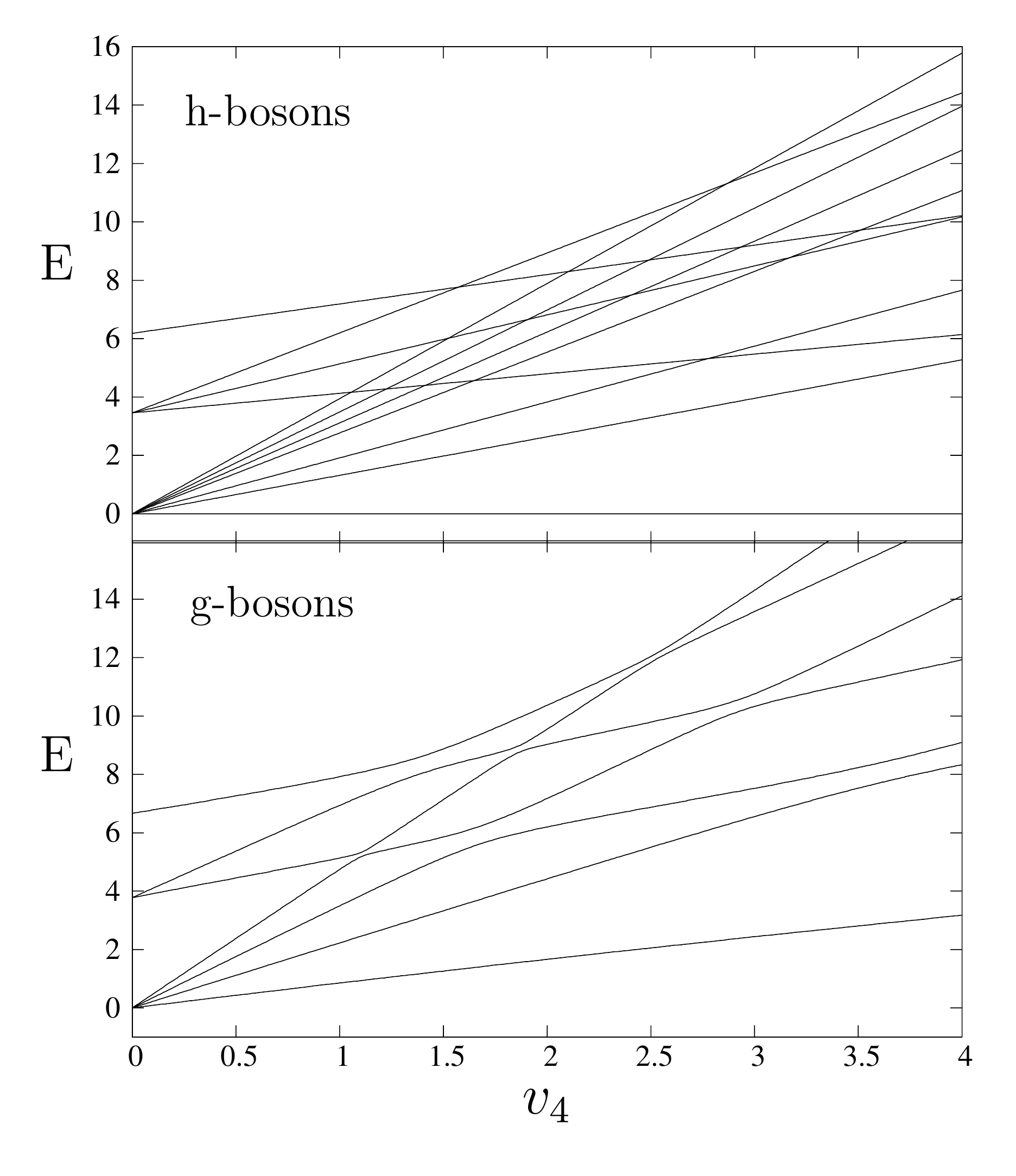}
\caption{The energy spectrum of six $h$ bosons (top)
or six $g$ bosons (bottom)
coupled to total angular momentum $J=2$
as a function of the interaction strength $\nu_4$.
For the $h$ bosons all crossings are unavoided
which is a consequence of the conservation of seniority.
In contrast, for $g$ bosons crossings are avoided
since there is no conservation of seniority.}
\label{f_gh}
\end{center}
\end{figure*}

From the general expression~(\ref{e_32cfp})
the following simple cases are obtained:
\begin{eqnarray}
[j^2(0)j;J|\}j^3[0]J]&=&
\sqrt{\frac{2j+1+2\sigma}{3(2j+1)}},
\nonumber\\[0mm]
[j^2(\lambda)j;J|\}j^3[0]J]&=&
\sigma\sqrt{\frac{4(2\lambda+1)}{3(2j+1)(2j+1+2\sigma)}},
\quad
\lambda\neq0,
\label{e_32cfpa}
\end{eqnarray}
where $\sigma\equiv(-)^{2j}$ is $+1$ for bosons and $-1$ for fermions.
This leads to the following condition (for $\lambda\neq0$)
valid for bosons and fermions:
\begin{eqnarray}
\delta_{\lambda I}+
2\sqrt{(2\lambda+1)(2I+1)}
\Biggl\{\begin{array}{ccc}j&j&\lambda\\j&j&I\end{array}\Biggr\}&=&
\frac{4\sqrt{(2\lambda+1)(2I+1)}}{(2j+1)(2j+1+2\sigma)}.
\label{e_con3c}
\end{eqnarray}
For seniority to be conserved by the interaction $\hat V_\lambda$,
this equation must be satisfied
for all even intermediate angular momenta $2\leq I\leq2p$.
Let us take $I=2$ and first consider $\lambda\neq I$.
For bosons the condition~(\ref{e_con3c}) then leads to the equation
\begin{eqnarray}
3\lambda^4
+6\lambda^3
-6[2j(j+1)-1]\lambda^2
&-&3[4j(j+1)-1]\lambda
\nonumber\\
&+&2j(j+1)(2j-1)(2j+1)=0.
\label{e_dio3b}
\end{eqnarray}
This should be considered
as a {\em diophantine} equation in $\lambda$
since only (positive, even) integer solutions in $\lambda$
have a physical meaning.
For $j=2$ the diophantine equation~(\ref{e_dio3b})
is satisfied for $\lambda=4$.
This confirms a known result namely
that {\em any} interaction between $d$ bosons is integrable
and conserves seniority, hence also $\hat V_4$.
More surprisingly, the equation
is also satisfied for $j=5$ and $\lambda=4$
and one may verify that in that case the condition~(\ref{e_con3c})
is equally valid for $I=4$, 6, 8 and 10.
This means that a $\hat V_4$ interaction
between $h$ bosons conserves seniority.
The result is illustrated in Fig.~\ref{f_gh}
where the $J=2$ spectrum of six $h$ bosons
is shown as a function of the interaction strength $\nu_4$.
To split states with different seniorities,
a constant (repulsive) pairing interaction $\hat V_0$ is taken
to which a variable $\hat V_4$ part is added.
The resulting hamiltonian is diagonalized numerically
with the code {\tt ArbModel}~\cite{Heinzeun}
which can compute the properties
of a system consisting of an arbitrary combination of bosons and/or fermions
interacting through two-body forces.
The figure confirms that there are no avoided crossings for $h$ bosons
since the only crossings that do occur
are between levels of different seniority
and those are unavoided.
For comparison, the $J=2$ spectrum of six $g$ bosons
is also shown as a function of the strength $\nu_4$,
and in this case the crossings are avoided.

To complete the analysis of the boson case,
for $\lambda=I=2$ the condition~(\ref{e_con3c}) leads to
\begin{equation}
8j^5+60j^4+50j^3-375j^2-373j+630=0.
\label{e_dio3ba}
\end{equation}
This equation has the integer solutions $j=1$ and $j=2$
as should be since any interaction
between $p$ or $d$ bosons is integrable
and conserves seniority.

For fermions the condition~(\ref{e_con3c})
leads to the diophantine equation
\begin{eqnarray}
3\lambda^4
+6\lambda^3
-6[2j(j+1)-1]\lambda^2
&-&3[4j(j+1)-1]\lambda
\nonumber\\
&+&2j(j+1)(2j+1)(2j+3)=0.
\label{e_dio3f}
\end{eqnarray}
This equation is satisfied
for $(j,\lambda)=(5/2,4)$, $(7/2,4)$ and $(7/2,6)$
which confirms the known result that for fermions with $j\leq7/2$
any interaction is diagonal in seniority~\cite{Shalit63,Talmi93}.
Finally, to complete the analysis for fermions,
for $\lambda=I=2$ the condition~(\ref{e_con3c}) leads to
\begin{equation}
8j^5-20j^4-110j^3+245j^2+327j-630=0,
\label{e_dio3fa}
\end{equation}
which has the half-odd-integer solutions $j=3/2$, 5/2 and 7/2,
again as should be.

As an amusing aside, note that
the diophantine equations~(\ref{e_dio3ba}) and (\ref{e_dio3fa})
also allow {\em negative} solutions, namely,
$-5/2$, $-7/2$ and $-9/2$ for the bosons and $-2$ and $-3$ for the fermions.
It thus transpires that the negative solutions $x$ for the particles of one statistics
correspond to the positive solutions $-x-1$ for the particles of the other statistics.

\subsection{Seniority conservation for a general interaction}
\label{ss_gen}
Let us next consider the condition of seniority conservation
for a general interaction
$\hat V=\sum_\lambda \nu_\lambda\hat V_\lambda$.
The analogue of the condition~(\ref{e_con3}) is
\begin{equation}
\frac{\langle j^3[0]J|\hat V|j^3[I]J\rangle}
{\langle j^3[0]J|\hat V|j^3[0]J\rangle}=
\langle j^3[0]J|j^3[I]J\rangle,
\label{e_con3gen}
\end{equation}
where it is again assumed that $J=j$ and $I\neq0$.
This leads to the following relation
between the coefficients $\nu_\lambda$:
\begin{eqnarray}
\sum_\lambda\;
[j^2(\lambda)j;J|\}j^3[0]J]&&\left(
[j^2(0)j;J|\}j^3[0]J]\;
[j^2(\lambda)j;J|\}j^3[I]J]\right.
\label{e_con3fin}\\
&&\left.\!\!-
[j^2(0)j;J|\}j^3[I]J]\;
[j^2(\lambda)j;J|\}j^3[0]J]\right)
\nu_\lambda=0.
\nonumber
\end{eqnarray}
With use of the explicit expressions for the various CFPs
this can be cast into the following form:
\begin{eqnarray}
&&\sum_{\lambda=2}^{2p}\sqrt{2\lambda+1}
\left(\delta_{\lambda I}+2\sqrt{(2\lambda+1)(2I+1)}
\Biggl\{\begin{array}{ccc}j&j&\lambda\\ j&j&I\end{array}\Biggr\}\right.
\nonumber\\
&&\qquad\qquad\qquad\qquad\qquad\qquad-\left.
\frac{4\sqrt{(2\lambda+1)(2I+1)}}{(2j+1)(2j+1+2\sigma)}
\right)\nu_\lambda=0.
\label{e_con3expl}
\end{eqnarray}
This condition has been derived previously in a variety of ways
mostly for fermions~\cite{Shalit63,Talmi93,Rowe01,Rosensteel03}.
The result~(\ref{e_con3expl}) shows that a simple expression exists
which covers both the boson and the fermion case.
Although Eq.~(\ref{e_con3expl}) determines all constraints
on the matrix elements $\nu_\lambda$
by varying $I$ between 2 and $2p$,
it does not tell us how many of those are independent.
This number turns out to be
$\lfloor j/3\rfloor$ for bosons and $\lfloor(2j-3)/6\rfloor$ for fermions,
the number of independent seniority $v=3$ states~\cite{Ginocchio93}.
Hence no condition on the matrix elements $\nu_\lambda$
follows for $j=1$ and 2, and for $j=3/2$, 5/2 and 7/2.
For higher values of $j$ one finds
\begin{eqnarray}
j=3&~:~&11\nu_2-18\nu_4+7\nu_6=0,
\nonumber\\
j=4&:&65\nu_2-30\nu_4-91\nu_6+56\nu_8=0,
\nonumber\\
j=5&:&3230\nu_2-2717\nu_6-3978\nu_8+3465\nu_{10}=0,
\nonumber\\
j=6&:&22610\nu_2+4788\nu_4-8099\nu_6-24106\nu_8-23793\nu_{10}+28600\nu_{12}=0,
\nonumber\\
&&90440\nu_2+156807\nu_4-409136\nu_6+290666\nu_8-275352\nu_{10}
\nonumber\\
&&\qquad+146575\nu_{12}=0,
\nonumber\\
&\vdots&\nonumber
\end{eqnarray}
and
\begin{eqnarray}
j=9/2&~:~&65\nu_2-315\nu_4+403\nu_6-153\nu_8=0,
\nonumber\\
j=11/2&:&1020\nu_2-3519\nu_4+637\nu_6+4403\nu_8-2541\nu_{10}=0,
\nonumber\\
j=13/2&:&1615\nu_2-4275\nu_4-1456\nu_6+3196\nu_8+5145\nu_{10}-4225\nu_{12}=0.
\nonumber\\
j=15/2&:&1330\nu_2-2835\nu_4-1807\nu_6+612\nu_8+3150\nu_{10}+3175\nu_{12}
\nonumber\\
&&\qquad-3625\nu_{14}=0,
\nonumber\\
&&77805\nu_2-169470\nu_4-85527\nu_6-4743\nu_8+222768\nu_{10}
\nonumber\\
&&\qquad+168025\nu_{12}-208858\nu_{14}=0.
\nonumber\\
&\vdots&\nonumber
\end{eqnarray}
for bosons and fermions, respectively.
The coefficient $\nu_0$ is absent from all equations
since pairing ($\lambda=0$) is known to conserve seniority.
Note that for $j=5$ there is no term in $\nu_4$
which is consistent with the results of Sect.~\ref{ss_dio}.

These results establish the necessary conditions
for an interaction to be seniority conserving
by imposing vanishing matrix elements
between seniority $v=1$ and $v=3$ three-particle states.
It can be shown with use of generic properties of CFPs
that these are also {\em sufficient} conditions~\cite{Shalit63,Talmi93}.
This means that the conditions~(\ref{e_con3expl})
are necessary and sufficient
for seniority to be a conserved quantum number
in a system of $N$ identical particles.

\section{Partial seniority conservation for four identical particles}
\label{s_sen4}
Let us now turn our attention to the four-particle case.
This analysis will, of course, confirm the results of the previous section
pertaining to seniority conservation in $N$-particle systems
but the particular interest of this section
concerns the possibility of {\em partial} seniority conservation.
It is important to clarify first
what is meant by partial dynamical symmetry
which is an enlargement of the concept of dynamical symmetry
as defined in Sect.~\ref{s_solv}.
The idea is to relax the conditions of {\em complete} solvability
and this can be done in essentially two different ways:
\begin{enumerate}
\item
{\it Some of the eigenstates keep all of the quantum numbers.}
In this case the properties of solvability, good quantum numbers,
and symmetry-dictated structure are fulfilled exactly,
but only by a subset of eigenstates~\cite{Alhassid92,Leviatan96,Garcia09}.
\item
{\it All eigenstates keep some of the quantum numbers.}
In this case eigenstates are not solvable,
yet some quantum numbers (of the conserved symmetries)
are retained.
In general, this type of partial dynamical symmetry arises
if the hamiltonian preserves some of the quantum numbers
in a dynamical-symmetry classification
while breaking others~\cite{Leviatan86,Isacker99}.
\end{enumerate}
Combinations of (1) and (2) are possible as well,
for example, if some of the eigenstates
keep some of the quantum numbers~\cite{Leviatan02}.

A further clarification concerning the notion of solvability is needed.
One might argue, for example,
that, as long as the hamiltonian matrix is of finite size
(as it is always the case in this paper),
its eigenvalues and eigenvectors can be determined
in a finite number of steps
and that as a consequence the secular equation is exactly solvable.
The condition of solvability adopted here is stronger,
and requires the property
of a predetermined structure of the eigenvector,
independent of the parameters in the hamiltonian.
So, an eigenstate will be called solvable only if its structure
is independent of the interaction matrix elements $\nu_\lambda$.
This is also the definition adopted by Talmi~\cite{Talmi10}
who showed that, if an eigenstate is solvable in this sense,
its energy is a linear combination of the $\nu_\lambda$
with coefficients that are rational non-negative numbers.

How do seniority-conserving interactions fit in this classification?
If the conditions~(\ref{e_con3expl}) are satisfied
by an interaction $\hat V$,
all its eigenstates carry the seniority quantum number $v$
and, consequently, the second type
of partial dynamical symmetry applies.
In addition, some of the eigenstates are completely solvable.
For example, the eigenstate with seniority $v=0$
of a seniority-conserving interaction
(this corresponds to the ground state of an even-even nucleus)
has a structure independent of the hamiltonian's parameters
and an analytic expression is available for its energy.
So, one concludes that seniority-conserving interactions
in general satisfy the second type of partial dynamical symmetry
but with the added feature that some states are completely solvable.

And what about more general interactions?
More specifically, is it possible to construct seniority-mixing interactions,
some of the eigenstates of which have good seniority?
An example was given by Escuderos and Zamick~\cite{Escuderos06}
who pointed out that four fermions in a $j=9/2$ shell
display one $J=4$ and one $J=6$ state 
both of which have seniority $v=4$
for an {\em arbitrary} interaction.
This is an example of a partial symmetry,
where seniority is broken for most but not for all states.

\subsection{Conditions for seniority conservation}
\label{ss_con4}
To shed light on the problem of partial seniority conservation,
the four-particle case can be analyzed,
displaying a close analogy with the three-particle case
reviewed in Sect.~\ref{s_sen3}.
A four-particle state can be written as $|j^2(R)j^2(R');J\rangle$
where two particles are first coupled to angular momentum $R$,
the next two particles to $R'$
and the intermediate angular momenta $R$ and $R'$ to total $J$.
This state is not (anti-)symmetric in all four particles
and can be made so by applying the (anti-)symmetry operator $\hat P$,
\begin{eqnarray}
|j^4[II']J\rangle
&\propto&{\hat P}|j^2(I)j^2(I');J\rangle
\nonumber\\
&=&\sum_{RR'}\;
[j^2(R)j^2(R');J|\}j^4[II']J]\;
|j^2(R)j^2(R');J\rangle,
\label{e_4pstate}
\end{eqnarray}
where $[j^2(R)j^2(R');J|\}j^4[II']J]$ is a four-to-two-particle CFP.
The notation in square brackets $[II']$ implies that the state~(\ref{e_4pstate})
is constructed from a parent
with intermediate angular momenta $I$ and $I'$.
It is implicitly assumed
that $I$ and $I'$ as well as $R$ and $R'$ are even.

The remarks made in the three-particle case
concerning non-orthogonality and over-completeness apply also here.
Because of the difficulties associated with a non-orthogonal basis,
it will sometimes be advantageous to convert to an orthogonal one,
which can be achieved through a standard Gram-Schmidt procedure.
Given an ordered set of $p$ non-orthogonal bases states,
\begin{equation}
|j^4[I_1I'_1]J\rangle,|j^4[I_2I'_2]J\rangle,\dots,|j^4[I_pI'_p]J\rangle,
\end{equation}
the orthonormalized bases states will be denoted as
\begin{eqnarray}
|j^4[\widetilde{I_1I'_1}]J\rangle&=&
\frac{1}{\sqrt{o_{11}}}|j^4[I_1I'_1]J\rangle,
\nonumber\\
|j^4[\widetilde{I_2I'_2}]J\rangle&=&
\frac{1}{\sqrt{o_{22}-(\tilde o_{21})^2}}
\left(|j^4[I_2I'_2]J\rangle-\tilde o_{21}|j^4[\widetilde{I_1I'_1}]J\rangle\right),
\nonumber\\&\vdots&
\nonumber\\
|j^4[\widetilde{I_kI'_k}]J\rangle&=&
\frac{1}{\sqrt{N_k}}
\left(|j^4[I_kI'_k]J\rangle-\sum_{i=1}^{k-1}\tilde o_{ki}|j^4[\widetilde{I_iI'_i}]J\rangle\right),
\nonumber\\&\vdots&
\label{e_gsbasis}
\end{eqnarray}
until $k=p$, with
\begin{equation}
N_k=o_{kk}-\sum_{i=1}^{k-1}(\tilde o_{ki})^2,
\qquad
\tilde o_{ki}=\langle j^4[I_kI'_k]J|j^4[\widetilde{I_iI'_i}]J\rangle.
\end{equation}

The four-to-two-particle CFP is known in closed form,
\begin{eqnarray}
&&[j^2(R)j^2(R');J|\}j^4[II']J]
\nonumber\\
&&\qquad=\frac{1}{\sqrt{{N}_{jJ}^{II'}}}
\left(\delta_{RI}\delta_{R'I'}+(-)^J\delta_{RI'}\delta_{R'I}
+4\sigma\left[\begin{array}{ccccc}
j&\;&j&\;&R\\ j&&j&&R' \\ I&&I'&&J
\end{array}\right]\right),
\label{e_42cfp}
\end{eqnarray}
where the symbol in square brackets
is related to the $9j$ symbol through
\begin{equation}
\left[\begin{array}{ccc}
j_1&j_2&J_{12}\\j_3&j_4&J_{34}\\J_{13}&J_{24}&J
\end{array}\right]=
\hat J_{12}\hat J_{34}\hat J_{13}\hat J_{24}
\left\{\begin{array}{ccc}
j_1&j_2&J_{12}\\j_3&j_4&J_{34}\\J_{13}&J_{24}&J
\end{array}\right\},
\label{e_9j}
\end{equation}
with $\hat J=\sqrt{2J+1}$.
With use of the following sum over {\em even} values of $R$ and $R'$,
\begin{eqnarray}
&&\sum_{RR'\;{\rm even}}
\left[\begin{array}{ccccc}
j&\;&j&\;&R\\ j&&j&&R' \\ I&&I'&&J
\end{array}\right]
\left[\begin{array}{ccccc}
j&\;&j&\;&R\\ j&&j&&R' \\ L&&L'&&J
\end{array}\right]
={\frac 1 4}
\left(\delta_{IL}\delta_{I'L'}
+(-)^{I+I'+J}\delta_{IL'}\delta_{I'L}
\phantom{\left[\begin{array}{c}j\\ j \\ L\end{array}\right]}
\right.
\nonumber\\
&&\qquad\qquad\qquad\qquad\quad\left.
+2\sigma\left((-)^{I+L}+(-)^{I'+L'}\right)
\left[\begin{array}{ccccc}
j&\;&j&\;&I\\ j&&j&&I' \\ L&&L'&&J
\end{array}\right]\right),
\label{e_sum9j}
\end{eqnarray}
the normalization coefficient can be obtained as
\begin{equation}
{N}_{jJ}^{II'}=
6\left(1+(-)^J\delta_{II'}+
4\sigma\left[\begin{array}{ccccc}
j&\;&j&\;&I\\ j&&j&&I' \\ I&&I'&&J
\end{array}\right]\right).
\label{e_42norm}
\end{equation}
Both the overlap matrix
and the matrix element of the operator $\hat V_\lambda$
can be expressed in terms of the CFPs,
\begin{eqnarray}
&&\langle j^4[II']J|j^4[LL']J\rangle
\nonumber\\
&&\qquad=\sum_{RR'}\;
[j^2(R)j^2(R');J|\}j^4[II']J]\;
[j^2(R)j^2(R');J|\}j^4[LL']J],
\label{e_overlap4s}
\end{eqnarray}
\begin{eqnarray}
&&\langle j^4[II']J|\hat V_\lambda|j^4[LL']J\rangle
\nonumber\\
&&\qquad=6\sum_R\;
[j^2(R)j^2(\lambda);J|\}j^4[II']J]\;
[j^2(R)j^2(\lambda);J|\}j^4[LL']J].
\label{e_hmat4}
\end{eqnarray}
With use of the result~(\ref{e_sum9j})
the first of these sums can be carried out,
yielding the expression
\begin{eqnarray}
&&\langle j^4[II']J|j^4[LL']J\rangle
\nonumber\\
&&\qquad=\frac{6}{\sqrt{{N}_{jJ}^{II'}{N}_{jJ}^{LL'}}}
\left(\delta_{IL}\delta_{I'L'}+(-)^J\delta_{IL'}\delta_{I'L}
+4\sigma\left[\begin{array}{ccccc}
j&\;&j&\;&I\\ j&&j&&I' \\ L&&L'&&J
\end{array}\right]\right)
\nonumber\\
&&\qquad=
\frac{6}{\sqrt{{N}_{jJ}^{II'}}}
[j^2(I)j^2(I');J|\}j^4[LL']J].
\label{e_overlap4}
\end{eqnarray}

The four-particle case with $J=0$
is equivalent to three particles coupled to $J=j$
which was considered in Sect.~\ref{s_sen3}.
Therefore it is assumed in the following that $J\neq0$,
corresponding to four-particle states with seniority $v=2$ or $v=4$.
By definition the seniority $v=2$ four-particle state is
\begin{equation}
|j^4,v=2,J\rangle=|j^4[0J]J\rangle.
\label{e_v2}
\end{equation}
A seniority $v=4$ state is orthogonal to this state
and can thus be written as
\begin{equation}
|j^4[II'],v=4,J\rangle=
|j^4[II']J\rangle-
\langle j^4[II']J|j^4[0J]J\rangle|j^4[0J]J\rangle.
\label{e_v4}
\end{equation}
There can be more than one seniority $v=4$ state for a given $J$
in which case the indices $[II']$ may serve as an additional label.
Seniority conservation of $\hat V_\lambda$ implies
$\langle j^4[0J]J|\hat V_\lambda|j^4[II'],v=4,J\rangle=0$
or
\begin{equation}
\frac{\langle j^4[0J]J|\hat V_\lambda|j^4[II']J\rangle}
{\langle j^4[0J]J|\hat V_\lambda|j^4[0J]J\rangle}=
\langle j^4[0J]J|j^4[II']J\rangle.
\label{e_con4}
\end{equation}
With use of the expressions~(\ref{e_hmat4}) and~(\ref{e_overlap4})
this condition reduces to
\begin{eqnarray}
&&\frac{\sum_R\;
[j^2(R)j^2(\lambda);J|\}j^4[II']J]\;
[j^2(R)j^2(\lambda);J|\}j^4[0J]J]}
{\sum_{R}\;
[j^2(R)j^2(\lambda);J|\}j^4[0J]J]\;
[j^2(R)j^2(\lambda);J|\}j^4[0J]J]}
\nonumber\\
&&\qquad=
\frac{6}{\sqrt{{N}_{jJ}^{0J}}}
[j^2(0)j^2(J);J|\}j^4[II']J].
\label{e_con4a}
\end{eqnarray}

In the same way as in the three-particle case
one can also derive the condition of seniority conservation
for a general interaction $\hat V=\sum_\lambda \nu_\lambda\hat V_\lambda$.
The condition
\begin{equation}
\frac{\langle j^4[0J]J|\hat V|j^4[II']J\rangle}
{\langle j^4[0J]J|\hat V|j^4[0J]J\rangle}=
\langle j^4[0J]J|j^4[II']J\rangle,
\label{e_con4gen}
\end{equation}
leads to the following equation:
\begin{eqnarray}
&&\sum_{R\lambda}\;
[j^2(R)j^2(\lambda);J|\}j^4[0J]J]
\nonumber\\
&&\qquad\times\left(
[j^2(J)j^2(0);J|\}j^4[0J]J]\;
[j^2(R)j^2(\lambda);J|\}j^4[II']J]\right.
\nonumber\\
&&\qquad\left.\;\;-
[j^2(J)j^2(0);J|\}j^4[II']J]\;
[j^2(R)j^2(\lambda);J|\}j^4[0J]J]\right)
\nu_\lambda=0.
\label{e_con4fin}
\end{eqnarray}
Note the formal equivalence of this condition
to the one obtained in the three-particle case, Eq.~(\ref{e_con3fin}).
Insertion of the values for the four-to-two-particle CFPs
yields exactly the same constraints
as those derived in the three-particle case.

\subsection{Partial seniority conservation}
\label{ss_sen4}
Let us now turn our attention to the problem of partial seniority conservation
and derive the conditions for an interaction $\hat V$
to have {\em some} four-particle eigenstates with good seniority.
Note that there are a number of  `trivial' examples of this.
For example, if the total angular momentum $J$ is odd,
a four-particle state cannot be of seniority $v=0$ or $v=2$
and must necessarily have seniority $v=4$.
Also, for $J>2p$ the four-particle state must be of seniority $v=4$.
These trivial cases are not of interest here
but rather the situation
where both $v=2$ and $v=4$ occur for the same $J$
and where a general interaction $\hat V$
mixes the $v=2$ state with a subset of the $v=4$ states
but not with all.
Let us denote such a special $v=4$ state
as $|j^4,v=4,{\rm s},J\rangle$
and expand it in terms of the basis $|j^4[II']J\rangle$ discussed previously,
\begin{equation}
|j^4,v=4,{\rm s},J\rangle=
\sum_{II'\in\wp}\eta_{II'}|j^4[II']J\rangle,
\label{e_v4spec}
\end{equation}
where the sum runs over $q$ linearly independent combinations $[II']$ in the set $\wp$,
as many as there are independent four-particle states with angular momentum $J$.
For this state to be an eigenstate of $\hat V$ it should satisfy
\begin{equation}
\hat V|j^4,v=4,{\rm s},J\rangle=
E|j^4,v=4,{\rm s},J\rangle,
\label{e_conpa}
\end{equation}
in addition to the condition of orthogonality to the $v=2$ state,
\begin{equation}
\langle j^4[0J]J|j^4,v=4,{\rm s},J\rangle=0.
\label{e_conpb}
\end{equation}
Let us now focus on bosons with $3\leq j\leq5$ or fermions with $9/2\leq j\leq13/2$.
In these cases a general interaction
can be written as a {\em single} component $\hat V_\lambda$
plus an interaction $\hat V'$ that conserves seniority.
The conditions~(\ref{e_conpa}) and~(\ref{e_conpb})
must therefore be checked for a single $\lambda$ component only,
which can be arbitrarily chosen.
Hence one arrives at the conditions
\begin{equation}
\sum_{II'\in\wp}\eta_{II'}
\langle j^4[LL']J|\hat V_\lambda|j^4[II']J\rangle=
E_\lambda
\sum_{II'\in\wp}\eta_{II'}
\langle j^4[LL']J|j^4[II']J\rangle,
\label{e_conpa2}
\end{equation}
for the different indices $[LL']$ in the set $\wp$, and
\begin{equation}
\sum_{II'\in\wp}\eta_{II'}
\langle j^4[0J]J|\hat V_\lambda|j^4[II']J\rangle=0.
\label{e_conpb2}
\end{equation}
There are $q+1$ unknowns:
the $q$ coefficients $\eta_{II'}$ and the energy $E_\lambda$.
Equations~(\ref{e_conpa2}) and~(\ref{e_conpb2}) are also $q+1$ in number
and, together with the appropriate normalization condition for the $\eta_{II'}$,
they define an overcomplete set of equations in $\{\eta_{II'},E_\lambda\}$,
not satisfied in general but possibly for special values of $j$ and $J$.
Furthermore, according to the preceding  discussion,
if these equations are satisfied for one $\lambda$,
they must be valid for all $\lambda$
and in each case the solution yields $E_\lambda$,
the eigenvalue of $\hat V_\lambda$.

A symbolic solution of Eqs.~(\ref{e_conpa2}) and~(\ref{e_conpb2})
(for general $j$ and $J$) is difficult to obtain
but, using the expressions derived previously for the various matrix elements,
it is straightforward to find a particular solution for given $j$ and $J$.
In this way the finding of Refs.~\cite{Escuderos06,Zamick07}
is confirmed, that is, Eqs.~(\ref{e_conpa2}) and~(\ref{e_conpb2})
have a solution for $j=9/2,J=4$ and for $j=9/2,J=6$.
The resulting solvable states are given by
\begin{eqnarray}
|(9/2)^4,v=4,{\rm s},J=4\rangle&=&
\sqrt{\frac{25500}{25591}}|(9/2)^4[\widetilde{22}]4\rangle-
\sqrt{\frac{91}{25591}}|(9/2)^4[\widetilde{24}]4\rangle,
\nonumber\\
|(9/2)^4,v=4,{\rm s},J=6\rangle&=&
\sqrt{\frac{27132}{27257}}|(9/2)^4[\widetilde{24}]6\rangle+
\sqrt{\frac{125}{27257}}|(9/2)^4[\widetilde{26}]6\rangle.
\nonumber\\
\label{e_j9wave}
\end{eqnarray}
These states are identical to those of Eq.~(8) of Ref.~\cite{Isacker08}
but written here in the Gram-Schmidt basis defined in Eq.~(\ref{e_gsbasis}).
For the definition of this basis one starts, for $J=4$,
from the non-orthogonal set $|(9/2)^4[I_kI'_k]4\rangle$
with $[I_kI'_k]=[04]$, [22] and [24].
The second and third states
obtained after the Gram-Schmidt orthonormalization
are orthogonal to $|(9/2)^4[04]4\rangle$,
and hence have by definition seniority $v=4$.
A similar argument is valid for $J=6$
where the non-orthogonal set $|(9/2)^4[I_kI'_k]6\rangle$
has $[I_kI'_k]=[06]$, [24] and [26].
Consequently, the states~(\ref{e_j9wave}) have seniority $v=4$.

Furthermore, for each choice of $\lambda$,
the solution of the Eqs.~(\ref{e_conpa2}) and (\ref{e_conpb2}) yields $E_\lambda$
and these can be used to derive the following energy expressions:
\begin{eqnarray}
E[(9/2)^4,v=4,{\rm s},J=4]&=&
\frac{68}{33}\nu_2+\nu_4+\frac{13}{15}\nu_6+\frac{114}{55}\nu_8,
\nonumber\\
E[(9/2)^4,v=4,{\rm s},J=6]&=&
\frac{19}{11}\nu_2+\frac{12}{13}\nu_4+\nu_6+\frac{336}{143}\nu_8.
\label{e_j9energ4}
\end{eqnarray}

The wave functions of the two states are pre-determined
and their energies are linear combinations of the $\nu_\lambda$
with coefficients that are rational non-negative numbers.
These results are valid for an {\em arbitrary} interaction among $j=9/2$ fermions.
According to the discussion of Ref.~\cite{Talmi10},
the states are solvable,
independent of whether the interaction conserves seniority or not.

These results are, in fact, rather surprising
as can be seen from the structure of the energy matrices
for a general interaction $\hat V$.
For the $(9/2)^4$ states with angular momentum $J=4$ one finds
\begin{equation}
\left[\begin{array}{ccccc}
E[(9/2)^4[04]4]&~~&
\displaystyle\frac{1}{495}\sqrt{\frac{14}{2119}}\Delta E_1&~~&
\displaystyle\frac{2}{429}\sqrt{\frac{170}{489}}\Delta E_1\\
\displaystyle\frac{1}{495}\sqrt{\frac{14}{2119}}\Delta E_1&~~&
E[(9/2)^4[22]4]&~~&
\displaystyle\frac{10}{5379}\sqrt{\frac{595}{39}}\Delta E_2\\
\displaystyle\frac{2}{429}\sqrt{\frac{170}{489}}\Delta E_1&~~&
\displaystyle\frac{10}{5379}\sqrt{\frac{595}{39}}\Delta E_2&~~&
E[(9/2)^4[24]4]\\
\end{array}\right],
\label{e_matrix4}
\end{equation}
where the diagonal elements are given by
\begin{eqnarray}
E[(9/2)^4[04]4]&=&
\frac{3}{5}\nu_0+\frac{67}{99}\nu_2+\frac{746}{715}\nu_4+\frac{1186}{495}\nu_6+\frac{918}{715}\nu_8,
\nonumber\\
E[(9/2)^4[22]4]&=&
\frac{33161}{16137}\nu_2+\frac{1800}{1793}\nu_4+\frac{70382}{80685}\nu_6+\frac{18547}{8965}\nu_8,
\nonumber\\
E[(9/2)^4[24]4]&=&
\frac{2584}{5379}\nu_2+\frac{48809}{23309}\nu_4+\frac{65809}{26895}\nu_6+\frac{114066}{116545}\nu_8,
\label{e_dia4}
\end{eqnarray}
while in the off-diagonal elements
the following combinations of interaction matrix elements $\nu_J$ occur:
\begin{eqnarray}
\Delta E_1&=&
-65\nu_2+315\nu_4-403\nu_6+153\nu_8,
\nonumber\\
\Delta E_2&=&
-13\nu_2+9\nu_4+13\nu_6-9\nu_8.
\label{e_offdia}
\end{eqnarray}
The basis which is used for constructing the energy matrix~(\ref{e_matrix4})
are the states $|(9/2)^4[I_kI'_k]4\rangle\rangle$ discussed above.
If the combination $\Delta E_1$ vanishes in the matrix~(\ref{e_matrix4}),
no mixing occurs between the seniority $v=2$ and $v=4$ states,
and seniority is a good quantum number for the three eigenstates,
in agreement with the discussion of Subsect.~\ref{ss_gen}.

For completeness, the corresponding expressions for $J=6$ are
\begin{equation}
\left[\begin{array}{ccccc}
E[(9/2)^4[06]6]&~~&
\displaystyle-\frac{1}{1287}\sqrt{\frac{5}{97}}\Delta E_1&~~&
\displaystyle\frac{2}{2145}\sqrt{\frac{2261}{291}}\Delta E_1\\
\displaystyle-\frac{1}{1287}\sqrt{\frac{5}{97}}\Delta E_1&~~&
E[(9/2)^4[24]6]&~~&
\displaystyle\frac{10}{5379}\sqrt{\frac{595}{39}}\Delta E_2\\
\displaystyle\frac{2}{2145}\sqrt{\frac{2261}{291}}\Delta E_1&~~&
\displaystyle\frac{10}{5379}\sqrt{\frac{595}{39}}\Delta E_2&~~&
E[(9/2)^4[26]6]\\
\end{array}\right],
\label{e_matrix6}
\end{equation}
with the diagonal elements
\begin{eqnarray}
E[(9/2)^4[06]6]&=&
\frac{3}{5}\nu_0+\frac{34}{99}\nu_2+\frac{1186}{715}\nu_4+\frac{658}{495}\nu_6+\frac{1479}{715}\nu_8,
\nonumber\\
E[(9/2)^4[24]6]&=&
\frac{33049}{19206}\nu_2+\frac{25733}{27742}\nu_4+\frac{19331}{19206}\nu_6+\frac{65059}{27742}\nu_8,
\nonumber\\
E[(9/2)^4[26]6]&=&
\frac{1007}{3201}\nu_2+\frac{26370}{13871}\nu_4+\frac{7723}{3201}\nu_6+\frac{19026}{13871}\nu_8,
\label{e_dia6}
\end{eqnarray}
while the same combinations~(\ref{e_offdia})
occur in the off-diagonal elements.

The energy matrices~(\ref{e_matrix4}) and~(\ref{e_matrix6}) are $3\times3$
and, generally, none of the off-diagonal elements vanishes.
The eigenvalues therefore are roots of a cubic equation
and one may expect them to be complicated algebraic expressions
in terms of the interaction matrix elements $\nu_J$.
Surprisingly, this is not the case
and, for each of the matrices,
one eigenenergy is particularly simple
and given by one of the expressions in Eq.~(\ref{e_j9energ4}).

From Eq.~(\ref{e_j9energ4})
the following difference between the excitation energies is derived:
\begin{eqnarray}
&&E_{\rm x}[(9/2)^4,v=4,{\rm s},J=6]-E_{\rm x}[(9/2)^4,v=4,{\rm s},J=4]
\nonumber\\
&&\qquad=
-\frac{1}{3}\nu_2-\frac{1}{13}\nu_4+\frac{2}{15}\nu_6+\frac{18}{65}\nu_8.
\label{e_j9ediff}
\end{eqnarray}
Since the sum of the coefficients of the matrix elements
in the expression~(\ref{e_j9ediff}) is zero,
one can make the replacement $\nu_\lambda\rightarrow \nu_\lambda-\nu_0$
to arrive at the result
\begin{eqnarray}
&&E_{\rm x}[(9/2)^4,v=4,{\rm s},J=6]-E_{\rm x}[(9/2)^4,v=4,{\rm s},J=4]
\nonumber\\
&&\qquad=-\frac{1}{3}E_{\rm x}[(9/2)^2,v=2,J=2]
-\frac{1}{13}E_{\rm x}[(9/2)^2,v=2,J=4]
\nonumber\\
&&\qquad\phantom{=}
+\frac{2}{15}E_{\rm x}[(9/2)^2,v=2,J=6]
+\frac{18}{65}E_{\rm x}[(9/2)^2,v=2,J=8],
\label{e_j92eexc}
\end{eqnarray}
associating the excitation energies
of the $J=2$, 4, 6 and 8, seniority $v=2$ states in the two-particle system
with those of the $J=4$ and 6,
seniority $v=4$ states in the four-particle system.

Another interaction-independent result that can be derived
concerns transition matrix elements between the two states.
For example, an electric quadrupole transition
between two states characterized
by the expansion coefficients $\eta_{I_{\rm i}I'_{\rm i}}$ and $\eta_{I_{\rm f}I'_{\rm f}}$,
as in Eq.~(\ref{e_v4spec}),
has the $B$(E2) value
\begin{eqnarray}
&&B({\rm E2};j^4\{\eta_{I_{\rm i}I'_{\rm i}}\}J_{\rm i}
\rightarrow j^4\{\eta_{I_{\rm f}I'_{\rm f}}\}J_{\rm f})
\label{e_v4be2}\\
&&=20(2J_{\rm f}+1)(2j+1)
B({\rm E}2;2^+_1\rightarrow0^+_1)
\nonumber\\
&&\times
\left[
\sum_{I_{\rm i}I'_{\rm i}}
\sum_{I_{\rm f}I'_{\rm f}}
\sum_{R_{\rm i}R_{\rm f}R'}
\eta_{I_{\rm i}I'_{\rm i}}\sqrt{2R_{\rm i}+1}
[j^2(R_{\rm i})j^2(R');J_{\rm i}|\}j^4[I_{\rm i}I'_{\rm i}]J_{\rm i}]
\Biggl\{\begin{array}{ccc}
J_{\rm i}&R_{\rm i}&R'\\
R_{\rm f}&J_{\rm f}&2\end{array}\Biggr\}
\right.
\nonumber\\
&&\left.\phantom{
\sum_{I_{\rm i}I'_{\rm i}}
\sum_{I_{\rm f}I'_{\rm f}}
\sum_{R_{\rm i}R_{\rm f}R'}}\;\times
\eta_{I_{\rm f}I'_{\rm f}}\sqrt{2R_{\rm f}+1}
[j^2(R_{\rm f})j^2(R');J_{\rm f}|\}j^4[I_{\rm f}I'_{\rm f}]J_{\rm f}]
\Biggl\{\begin{array}{ccc}
R_{\rm i}&j&j\\
j&R_{\rm f}&2\end{array}\Biggr\}\right]^2,
\nonumber
\end{eqnarray}
where $B({\rm E}2;2^+_1\rightarrow0^+_1)$
is the $B$(E2) value in the two-particle system $j^2$.
For the two solvable states in the $j=9/2$ four-particle system
this reduces to the relation
\begin{eqnarray}
&&B({\rm E}2;(9/2)^4,v=4,{\rm s},J=6\rightarrow(9/2)^4,v=4,{\rm s},J=4)
\nonumber\\
&&\qquad=\frac{209475}{176468}B({\rm E}2;2^+_1\rightarrow0^+_1)
\approx1.19~B({\rm E}2;2^+_1\rightarrow0^+_1),
\label{e_j9be2}
\end{eqnarray}
which defines an interaction-independent relation
between the properties of the two- and four-particle systems.

We have searched for other examples of partial seniority conservation
but failed to find any for half-odd-integer values $j\neq9/2$.
So it transpires that the two solvable seniority $v=4$ states
of the four-particle system in the $j=9/2$ shell are unique.
The situation for bosons is different,
as discussed in Sect.~\ref{s_senbos}.

Although the mathematical derivation
of the necessary conditions for the existence of partial seniority conservation is clear,
a simple, intuitive reason for it is still lacking.
In Ref.~\cite{Zamick08} some progress
towards this goal has been made,
and in particular a partial understanding with analytic arguments
of the coefficients entering the energy expressions~(\ref{e_j9energ4})
has been achieved.
So far, the best explanation
of the anomalous partial conservation of seniority in the $j=9/2$ shell
has been given by Qi~\cite{Qi11}
who found an analytic derivation of this property
based on the uniqueness of the $|j^5,v=5,J=j\rangle$ state for $j=9/2$,
state which is no longer unique for $j>9/2$.

\section{Partial seniority conservation for $N$ identical bosons}
\label{s_senbos}
In Sect.~\ref{s_solv} are given
the sufficient conditions~(\ref{e_solv}) for a hamiltonian
describing a system of interacting bosons or fermions
to have a dynamical symmetry.
The derivation is based on a simple counting argument
which for bosons is as follows.
For identical bosons
the number of independent quadratic Casimir operators
in the canonical classification~(\ref{e_clasb})
is two for $\ell=1$ and three for $\ell>1$.
(In this section the notation $\ell$ instead of $j$
is used for the spin of the particles,
to emphasize that they are bosons.)
This matches the number of two-body interactions
for $p$ and $d$ bosons
which therefore are solvable systems.
If the spin of the bosons exceeds $\ell=2$
($f$ bosons and beyond),
there are more two-body interactions
than quadratic Casimir operators,
and a general two-body hamiltonian
does not have a dynamical symmetry.
In this section it is shown that
in systems of interacting bosons with spin $\ell\geq3$
many states occur to which additional constraints apply
as a consequence of which they are solvable.

For the present discussion it is convenient
to replace the quadratic Casimir operators in the hamiltonian~(\ref{e_hamb})
with equivalent operators
that are of pure two-body character.
The quadratic Casimir operator of U($n$) (with $n\equiv2\ell+1$) is,
up to a term linear in the boson number operator $\hat N$,
equivalent to a constant interaction between the bosons
which shall be denoted as $\hat C$.
The quadratic Casimir operator of SO($n$) is,
up to linear and quadratic terms in $\hat N$,
equivalent to a pairing interaction $\hat P$.
Finally, the quadratic Casimir operator of SO(3)
is identical to the square of the angular momentum operator $\hat J^2$;
again it is more convenient to retain only its two-body part
which shall be denoted as $\hat J_{\rm tb}^2$.
In terms of the earlier defined two-body operators $\hat V_\lambda$,
one has the identities
\begin{equation}
\hat C=\sum_\lambda\hat V_\lambda,
\qquad
\hat P=\hat V_0,
\qquad
\hat J_{\rm tb}^2=
\sum_\lambda[\lambda(\lambda+1)-2\ell(\ell+1)]\hat V_\lambda.
\label{e_ops}
\end{equation}
Given their connection with the quadratic Casimir operators
in the classification~(\ref{e_clasb}),
matrix elements of the operators $\hat C$, $\hat P$ and $\hat J_{\rm tb}^2$
can be found in closed form,
\begin{eqnarray}
\langle\ell^Nv\alpha J|\hat C|\ell^Nv\alpha J\rangle&=&
{\frac 1 2}N(N-1),
\nonumber\\
\langle\ell^Nv\alpha J|\hat P|\ell^Nv\alpha J\rangle&=&
N(N+n-2)-v(v+n-2),
\nonumber\\
\langle\ell^Nv\alpha J|\hat J_{\rm tb}^2|\ell^Nv\alpha J\rangle&=&
J(J+1)-N\ell(\ell+1).
\label{e_matops}
\end{eqnarray}

Let us consider now a state $|\ell^Nv\alpha J\rangle$
and ask the question whether values of boson number $N$,
seniority $v$, multiplicity label $\alpha$ and angular momentum $J$ exist
for which this is an eigenstate of an arbitrary two-body hamiltonian
$\hat V=\sum_\lambda\nu_\lambda\hat V_\lambda$
with analytic eigenvalues of the form
\begin{equation}
E(\ell^Nv\alpha J)=\sum_\lambda
a_\lambda\nu_\lambda,
\label{e_anal}
\end{equation}
in terms of the two-body matrix elements $\nu_\lambda$
with coefficients $a_\lambda$
that are functions of $\ell$, $N$, $v$, $\alpha$ and $J$.
The analysis concerns states of maximum seniority, $v=N$.
It has been remarked earlier that partial conservation of seniority
occurs `trivially' in a number of cases.
The most obvious example is
an $N$-boson state of stretched angular momentum $J=\ell N$.
It is clear that this state must have seniority $v=\ell$
and that only the stretched interaction matrix element with $\lambda=2\ell$
can contribute to its energy.
Formally, this result is obtained
from the expression for the matrix element
in terms of $N$-to-$(N-2)$-particle CFPs,
\begin{eqnarray}
\langle\ell^Nv\alpha J|\hat V_\lambda|\ell^Nv'\alpha'J\rangle&=&
\frac{N(N-1)}{2}
\sum_{v_1\alpha_1J_1}
[\ell^{N-2}(v_1\alpha_1J_1)\ell^2(\lambda);J|\}\ell^Nv\alpha J]
\nonumber\\&&\qquad\qquad\quad\times
[\ell^{N-2}(v_1\alpha_1J_1)\ell^2(\lambda);J|\}\ell^Nv'\alpha'J].
\label{e_twobody}
\end{eqnarray}
For a state with $J=\ell N$
the intermediate state is unique with $v_1=N-2$ and $J_1=\ell(N-2)$,
and the only interaction that couples $J_1$ to $J$ has $\lambda=2\ell$.
Since the corresponding CFP is unique it equals one
and hence
\begin{equation}
E(\ell^N,v=N,J=\ell N)=
\frac{N(N-1)}{2}\nu_{2\ell}.
\label{e_e3n}
\end{equation}

This argument clearly is only appropriate for $J=\ell N$ and $v=N$.
Similar but modified versions of it are possible for $J<\ell N$
and rely on the knowledge of the multiplicity $d^{(\ell)}_v(J)$,
which specifies how many times the angular momentum $J$
occurs for a given seniority $v$.
A closed formula is available for $d^{(\ell)}_v(J)$
in terms of an integral over characters of the orthogonal algebras SO($n$) and SO(3),
known from Weyl~\cite{Weyl39}.
This leads to the following complex integral~\cite{Gheorghe04}:
\begin{equation}
d^{(\ell)}_v(J)=
\frac{i}{2\pi}
\oint_{|z|=1}
\frac{(z^{2J+1}-1)(z^{2v+2\ell-1}-1)\prod_{k=1}^{2\ell-2}(z^{v+k}-1)}
{z^{\ell v+J+2}\prod_{k=1}^{2\ell-2}(z^{k+1}-1)}dz.
\label{e_mult}
\end{equation}
By virtue of Cauchy's theorem the multiplicity $d^{(\ell)}_v(J)$
is obtained as the negative of the residue of the integrand in Eq.~(\ref{e_mult}).

To illustrate how multiplicity enters
into the discussion of partial conservation of seniority,
it is easier to specify a value for the boson spin $\ell$.
Let us choose $\ell=3$.
Multiplicities for $f$ bosons
are given in Table~\ref{t_mult} up to seniority $v=15$.
\begin{table}
\caption{Multiplicity $d^{(f)}_v(J)$ for $f$ bosons up to seniority $v=15$.}
\label{t_mult}
\vspace{1ex}
\begin{tabular}{r|rrrrrrrrrrrrrrr}
\hline
$J\backslash v$& 1& 2& 3& 4& 5& 6& 7& 8& 9& 10& 11& 12& 13& 14& 15\\
\hline
0& 0& 0& 0& 1& 0& 1& 0& 1& 0& 2& 0& 2& 0& 2& 1\\[-2ex]
1& 0& 0& 1& 0& 1& 0& 2& 1& 2& 1& 3& 2& 4& 2& 5\\[-2ex]
2& 0& 1& 0& 1& 1& 2& 1& 3& 2& 4& 3& 5& 4& 7& 5 \\[-2ex]
3& 1& 0& 1& 1& 2& 2& 3& 2& 5& 4& 6& 5& 8& 7& 10 \\[-2ex]
4& 0& 1& 1& 2& 1& 3& 3& 5& 4& 6& 6& 9& 8& 11& 10 \\[-2ex]
5& 0& 0& 1& 1& 3& 2& 4& 4& 6& 6& 9& 8& 11& 11& 15 \\[-2ex]
6& 0& 1& 1& 2& 2& 4& 4& 6& 6& 9& 8& 12& 12& 15& 15 \\[-2ex]
7& 0& 0& 1& 1& 3& 3& 5& 5& 8& 8& 11& 11& 15& 15& 19 \\[-2ex]
8& 0& 0& 0& 2& 2& 4& 4& 7& 7& 10& 11& 14& 14& 19& 19 \\[-2ex]
9& 0& 0& 1& 1& 2& 3& 6& 6& 9& 9& 13& 14& 18& 18& 23 \\[-2ex]
10& 0& 0& 0& 1& 2& 4& 4& 7& 8& 12& 12& 16& 17& 22& 23 \\[-2ex]
11& 0& 0& 0& 0& 2& 2& 5& 6& 9& 10& 14& 15& 20& 21& 26 \\[-2ex]
12& 0& 0& 0& 1& 1& 3& 4& 7& 8& 12& 13& 18& 19& 24& 26 \\[-2ex]
13& 0& 0& 0& 0& 1& 2& 4& 5& 9& 10& 15& 16& 21& 23& 29 \\[-2ex]
14& 0& 0& 0& 0& 0& 2& 3& 6& 7& 11& 13& 18& 20& 26& 27 \\[-2ex]
15& 0& 0& 0& 0& 1& 1& 3& 4& 8& 10& 14& 16& 22& 24& 31 \\[-2ex]
16& 0& 0& 0& 0& 0& 1& 2& 5& 6& 10& 12& 18& 20& 26& 29 \\[-2ex]
17& 0& 0& 0& 0& 0& 0& 2& 3& 6& 8& 13& 15& 21& 24& 31 \\[-2ex]
18& 0& 0& 0& 0& 0& 1& 1& 3& 5& 9& 11& 16& 19& 26& 29 \\[-2ex]
19& 0& 0& 0& 0& 0& 0& 1& 2& 5& 6& 11& 14& 20& 23& 30 \\[-2ex]
20& 0& 0& 0& 0& 0& 0& 0& 2& 3& 7& 9& 14& 17& 24& 28 \\[-2ex]
21& 0& 0& 0& 0& 0& 0& 1& 1& 3& 5& 9& 12& 18& 21& 29 \\[-2ex]
22& 0& 0& 0& 0& 0& 0& 0& 1& 2& 5& 7& 12& 15& 22& 26 \\[-2ex]
23& 0& 0& 0& 0& 0& 0& 0& 0& 2& 3& 7& 9& 15& 19& 26 \\[-2ex]
24& 0& 0& 0& 0& 0& 0& 0& 1& 1& 3& 5& 10& 13& 19& 23 \\[-2ex]
25& 0& 0& 0& 0& 0& 0& 0& 0& 1& 2& 5& 7& 12& 16& 24 \\[-2ex]
26& 0& 0& 0& 0& 0& 0& 0& 0& 0& 2& 3& 7& 10& 16& 20 \\[-2ex]
27& 0& 0& 0& 0& 0& 0& 0& 0& 1& 1& 3& 5& 10& 13& 20 \\[-2ex]
28& 0& 0& 0& 0& 0& 0& 0& 0& 0& 1& 2& 5& 7& 13& 17 \\[-2ex]
29& 0& 0& 0& 0& 0& 0& 0& 0& 0& 0& 2& 3& 7& 10& 16 \\[-2ex]
30& 0& 0& 0& 0& 0& 0& 0& 0& 0& 1& 1& 3& 5& 10& 14 \\[-2ex]
31& 0& 0& 0& 0& 0& 0& 0& 0& 0& 0& 1& 2& 5& 7& 13 \\[-2ex]
32& 0& 0& 0& 0& 0& 0& 0& 0& 0& 0& 0& 2& 3& 7& 10 \\[-2ex]
33& 0& 0& 0& 0& 0& 0& 0& 0& 0& 0& 1& 1& 3& 5& 10 \\[-2ex]
34& 0& 0& 0& 0& 0& 0& 0& 0& 0& 0& 0& 1& 2& 5& 7 \\[-2ex]
35& 0& 0& 0& 0& 0& 0& 0& 0& 0& 0& 0& 0& 2& 3& 7 \\[-2ex]
36& 0& 0& 0& 0& 0& 0& 0& 0& 0& 0& 0& 1& 1& 3& 5 \\[-2ex]
37& 0& 0& 0& 0& 0& 0& 0& 0& 0& 0& 0& 0& 1& 2& 5 \\[-2ex]
38& 0& 0& 0& 0& 0& 0& 0& 0& 0& 0& 0& 0& 0& 2& 3 \\[-2ex]
39& 0& 0& 0& 0& 0& 0& 0& 0& 0& 0& 0& 0& 1& 1& 3 \\[-2ex]
40& 0& 0& 0& 0& 0& 0& 0& 0& 0& 0& 0& 0& 0& 1& 2 \\[-2ex]
41& 0& 0& 0& 0& 0& 0& 0& 0& 0& 0& 0& 0& 0& 0& 2 \\[-2ex]
42& 0& 0& 0& 0& 0& 0& 0& 0& 0& 0& 0& 0& 0& 1& 1 \\[-2ex]
43& 0& 0& 0& 0& 0& 0& 0& 0& 0& 0& 0& 0& 0& 0& 1 \\[-2ex]
44& 0& 0& 0& 0& 0& 0& 0& 0& 0& 0& 0& 0& 0& 0& 0 \\[-2ex]
45& 0& 0& 0& 0& 0& 0& 0& 0& 0& 0& 0& 0& 0& 0& 1\\
\hline
\end{tabular}
\end{table}
No state exists with angular momentum $J=3N-1$
and the highest possible, non-stretched angular momentum is $J=3N-2$.
This state is unique and must have seniority $v=N$
since $J=3N-2$ does not occur for lower seniorities,
that is, $d^{(f)}_v(3N-2)=0$ for $v=N-2,N-4,\dots$.
The eigenvalue of this state can be found
by noting from the expression~(\ref{e_twobody}) that
the interaction $\hat V_2$ cannot contribute to its energy
since the highest angular momentum
of the intermediate $f^{N-2}$ system is $J_1=3(N-2)$
which cannot couple with $\lambda=2$ to $J=3N-2$.
Hence one establishes the equations
\begin{eqnarray}
a_0=a_2&=&0,
\nonumber\\
a_0+a_2+a_4+a_6&=&
{\frac 1 2}N(N-1),
\nonumber\\
-24a_0-18a_2-4a_4+18a_6&=&(3N-2)(3N-1)-12N,
\label{e_c3n-2}
\end{eqnarray}
which can be solved to yield the energy expression
\begin{equation}
E(f^N,v=N,J=3N-2)=
\frac{6N-1}{11}\nu_4+
\frac{11N^2-23N+2}{22}\nu_6.
\label{e_e3n-2}
\end{equation}
The next highest angular momentum $J=3N-3$
is also unique and exists for $N\geq3$.
A closed energy expression can be found with the same argument,
\begin{equation}
E(f^N,v=N,J=3N-3)=
\frac{9N-3}{11}\nu_4+
\frac{11N^2-29N+6}{22}\nu_6.
\label{e_e3n-3}
\end{equation}

For angular momentum $J=3N-4$
one encounters the first case with multiplicity 2
(provided $N\geq4$).
The seniority of this state is still necessarily $v=N$
since $d^{(f)}_v(3N-4)=0$ for $v=N-2,N-4,\dots$.
However, unlike the previous cases,
it can couple with the interaction $\hat V_2$
to the stretched state of the intermediate $f^{N-2}$ system
with angular momentum $J_1=3(N-2)$,
so the interaction energy associated with $\hat V_2$
does not necessarily vanish but is given by
\begin{eqnarray}
&&\langle f^N,v=N,\alpha,J=3N-4|\hat V_2|f^N,v=N,\alpha,J=3N-4\rangle
\nonumber\\&&\qquad=
\frac{N(N-1)}{2}
[f^{N-2}(J_1=3N-6)f^2(2);J|\}f^Nv\alpha J]^2.
\end{eqnarray}
There are {\em two} states with $J=3N-4$ and $v=N$,
characterized by $\alpha_1$ and $\alpha_2$,
and only {\em one} intermediate state with $J_1=3N-6$.
One can therefore always choose a linear combination
of  $\alpha_1$ and $\alpha_2$, say $\bar\alpha$,
such that
\begin{equation}
[f^{N-2}(J_1=3N-6)f^2(2);J|\}f^Nv\bar\alpha J]=0.
\end{equation}
This state satisfies
\begin{equation}
\langle f^N,v=N,\alpha_i,J=3N-4|\hat V_2|f^N,v=N,\bar\alpha,J=3N-4\rangle=0,
\end{equation}
and by the same argument as before
one can derive its energy in closed form,
\begin{equation}
E(f^N,v=N,\bar\alpha,J=3N-4)=
\frac{12N-6}{11}\nu_4+
\frac{11N^2-35N+12}{22}\nu_6.
\label{e_e3n-4}
\end{equation}
The next highest angular momentum $J=3N-5$
has also multiplicity 2 for $N\geq5$.
A closed energy expression can be found with the same argument,
\begin{equation}
E(f^N,v=N,\bar\alpha,J=3N-5)=
\frac{15N-10}{11}\nu_4+
\frac{11N^2-41N+20}{22}\nu_6.
\label{e_e3n-5}
\end{equation}

The next case with angular momentum $J=3N-6$
presents the additional complication
that its seniority quantum number is not unique
but can be $v=N$ or $v=N-2$.
It can still be dealt with in the following way.
There are {\em two} intermediate states
in the expression for the matrix element of $\hat V_2$,
\begin{eqnarray}
&&\langle f^N,v=N,\alpha,J=3N-6|\hat V_2|f^N,v=N,\alpha,J=3N-6\rangle
\nonumber\\&&\qquad=
\frac{N(N-1)}{2}
\sum_{J_1}
[f^{N-2}(J_1)f^2(2);J|\}f^Nv\alpha J]^2,
\end{eqnarray}
with $J_1=3N-6$ or $3N-8$.
However, the state with angular momentum $J=3N-6$
has multiplicity 3 (for $N\geq6$)
and hence a linear combination $\bar\alpha$ can always be chosen
such that
\begin{equation}
[f^{N-2}(J_1)f^2(2);J|\}f^Nv\bar\alpha J]=0,
\qquad{\rm for}\quad
J_1=3N-6, 3N-8.
\label{e_cfp3n-6}
\end{equation}
As a consequence, the contribution of $\hat V_2$
to the energy of the state $|f^N,v=N,\bar\alpha,J=3N-6\rangle$ vanishes.
In addition, one has from Eq.~(\ref{e_twobody})
and the vanishing CFPs~(\ref {e_cfp3n-6}) that
\begin{equation}
\langle f^N,v=N-2,J=3N-6|\hat V_2|f^N,v=N,\bar\alpha,J=3N-6\rangle=0,
\end{equation}
that is, the state does not mix with the state with seniority $v=N-2$.
This is thus a first example of `non-trivial' conservation of seniority
since one has for a given angular momentum $J$ several possible seniorities $v$
and one state with seniority $v=N$
that does not mix with states of lower seniority.
Its eigenvalue is found to be
\begin{equation}
E(f^N,v=N,\bar\alpha,J=3N-6)=
\frac{18N-15}{11}\nu_4+
\frac{11N^2-47N+30}{22}\nu_6.
\label{e_e3n-6}
\end{equation}

The arguments for finding states which conserve seniority
while all others do not,
become increasingly complex
but are still valid for the next two cases
with angular momentum $J=3N-7$ ($N\geq7$)
and $J=3N-8$ ($N\geq8$).
The eigenvalue expressions are
\begin{equation}
E(f^N,v=N,\bar\alpha,J=3N-7)=
\frac{21N-21}{11}\nu_4+
\frac{11N^2-53N+42}{22}\nu_6,
\label{e_e3n-7}
\end{equation}
and
\begin{equation}
E(f^N,v=N,\bar\alpha,J=3N-8)=
\frac{24N-28}{11}\nu_4+
\frac{11N^2-59N+56}{22}\nu_6.
\label{e_e3n-8}
\end{equation}

The preceding analysis can be generalized to bosons with any spin $\ell$.
The solvable states satisfy the following energy expression:
\begin{equation}
E(\ell^N,v=N,\bar\alpha,J=\ell N-q)=
a^\ell_{Nq}\,\nu_{2\ell-2}+b^\ell_{Nq}\,\nu_{2\ell},
\label{e_elnq}
\end{equation}
with
\begin{eqnarray}
a^\ell_{Nq}&=&\frac{q\ell N-(q-1)q}{4\ell-1},
\nonumber\\
b^\ell_{Nq}&=&\frac{(4\ell-1)N^2-[(4\ell-1)+2q\ell]N+(q-1)q}{2(4\ell-1)}.
\label{e_elnq}
\end{eqnarray}
According to this analysis,
partial conservation of seniority
is restricted to states with maximal seniority, $v=N$,
and occurs at the high end of allowed angular momenta.
The surprising aspect of the result~(\ref{e_elnq})
is that partial conservation of seniority persists
for high values of $q$, up to $q=8$.

\section{Applications in fermionic systems}
\label{s_appfer}
Many studies exist which show the relevance of seniority in nuclei.
It is not the intention here to review all such applications,
many of which are discussed in Ref.~\cite{Talmi93}.
Rather, two consequences
of the {\em partial} conservation of seniority are pointed out,
which are related to the existence of seniority isomers
and to properties of one-nucleon transfer.

\subsection{Seniority isomers}
\label{ss_isomer}
Isomers are metastable quantum states.
In nuclei isomers generally adopt a configuration
which is different from those of states at lower energy
and their decay is therefore hindered.
The type of configuration change determines the nature of the isomer
and hence one distinguishes, for example,
shape isomers, spin isomers and $K$ isomers~\cite{Walker99}.

Seniority isomers exist by virtue of seniority and its associated selection rules.
Generally, and in particular in semi-magic nuclei,
states with low seniority occur at low energy.
For example, the ground state of an even-even semi-magic nucleus
has approximately seniority $v\approx0$
(all nucleons in pairs coupled to $J=0$)
while its yrast levels with angular momenta $J=2,4,6,\dots$
have seniority $v\approx2$ (containing one `broken' pair with $J\neq0$).
Seniority isomerism is expected to occur in semi-magic nuclei
because electric quadrupole (E2) transitions between states with seniority $v=2$
are small when the valence shell is close to half-filled.
This result is a consequence of the fact that
the matrix elements of even tensor operators---and
hence also of the quadrupole operator---between
states with seniority $v=2$ vanish at mid-shell~\cite{Shalit63,Talmi93}.

Examples of seniority isomers have been found
in the $N=50$ isotones
with protons dominantly confined to the $\pi1g_{9/2}$ shell~\cite{Jaklevic69}.
In particular, the $J^\pi=8^+$ levels
in $^{92}$Mo ($Z=42$), $^{94}$Ru ($Z=44$), $^{96}$Pd ($Z=46$) and $^{98}$Cd ($Z=48$)
have half-lives of 0.190(3), 71(4), 2.10(21) and 0.48(16)~$\mu$s, respectively,
resulting from a combination of slow E2 decay
and a small energy difference with the $J^\pi=6^+$ level below it.
A review is given by Grawe {\it et al.}~\cite{Grawe97}.
On the basis of similar arguments
one would expect the same phenomenon to occur
in the neutron-rich nickel ($Z=28$) isotopes from $^{70}$Ni to $^{76}$Ni
with neutrons dominantly confined to the $\nu1g_{9/2}$ shell.
Isomers with $J^\pi=8^+$ are indeed observed in $^{70}$Ni and $^{76}$Ni
with half-lives of 0.232(1) and $0.59^{+18}_{-11}$~$\mu$s, respectively,
but, in spite of intensive searches~\cite{Sawicka03},
none was found so far in $^{72}$Ni or $^{74}$Ni.
Only recently, a possible $J^\pi=(8^+)$ level
was identified in $^{72}$Ni at an excitation energy 2590~keV,
decaying to a $J^\pi=(6^+)$ level with the emission of a 199-keV gamma~\cite{Chiara11};
only an upper limit of 20~ns could be determined for the half-life
which therefore seems to exclude the isomeric character of this level.

An explanation of these observations is given in this subsection.
The problem was discussed by Grawe {\it et al.}~\cite{Grawe02}
who noted that the disappearance of the isomers in $^{72}$Ni and $^{74}$Ni
is related to an inversion of levels with seniority $v=2$ and $v=4$.
It will be shown that an analytic explanation exists
on the basis of the results derived in Sect.~\ref{s_sen4}.

\begin{figure}
\begin{center}
\includegraphics[width=13.8cm]{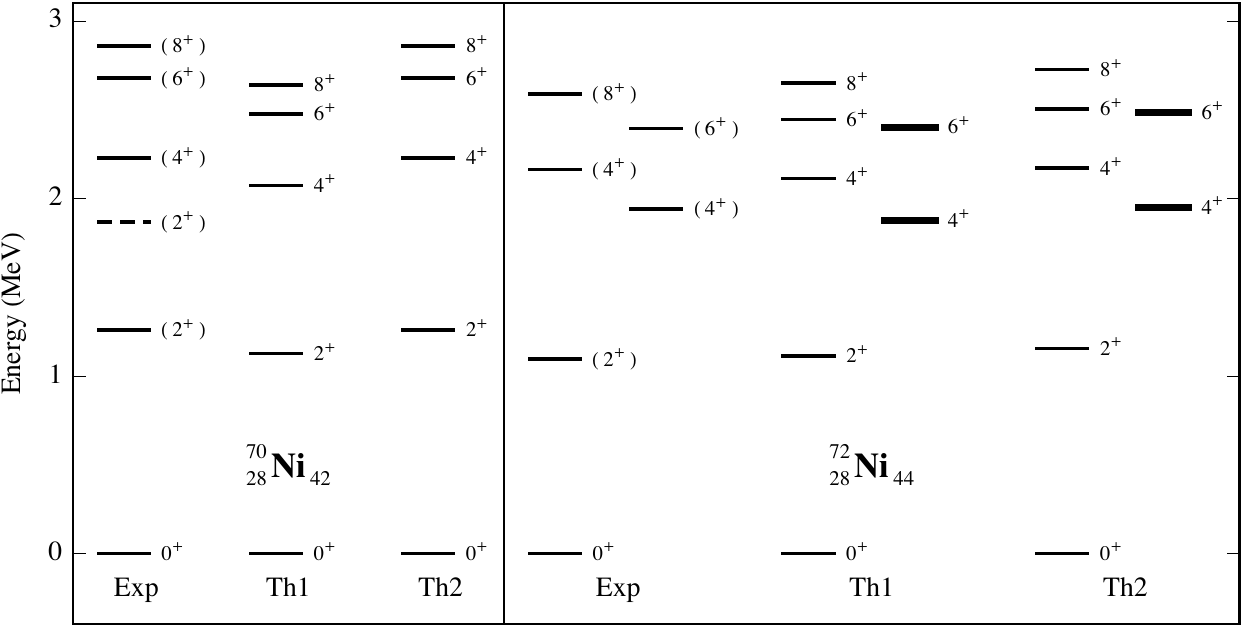}
\includegraphics[width=13.8cm]{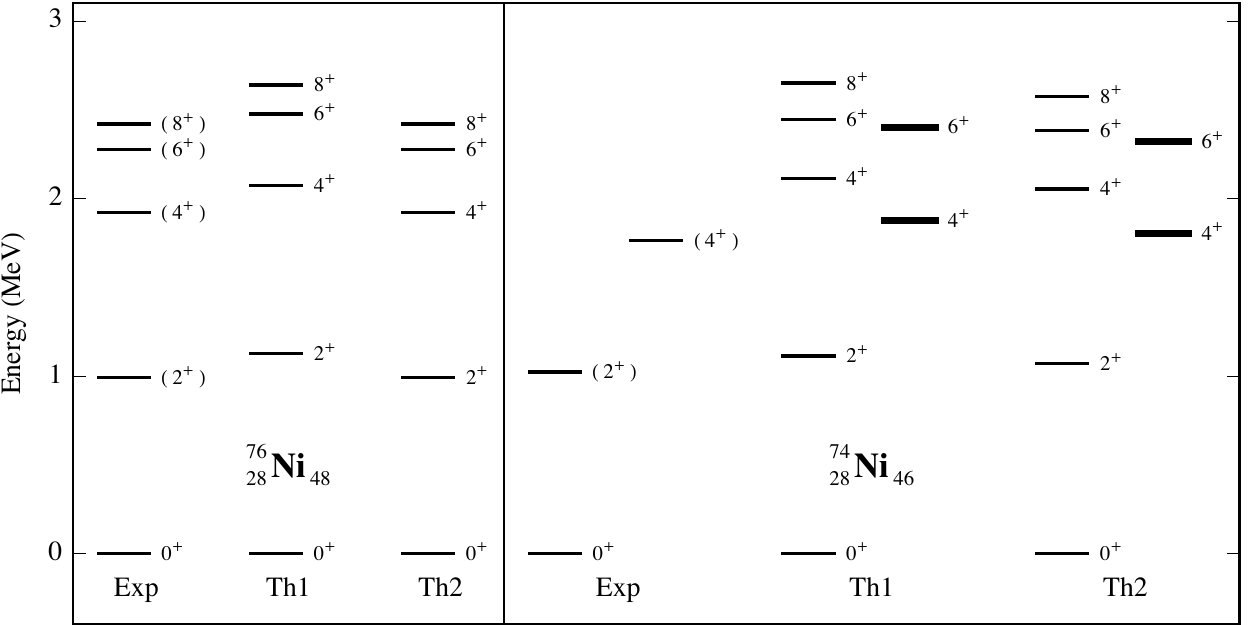}
\caption{The low-energy spectra of the nickel isotopes $^{70-76}$Ni.
The left-hand columns `Exp' show the observed levels
while the columns `Th1' and `Th2'
contain the results of a $(\nu1g_{9/2})^n$ shell-model calculation
with constant or linearly varying two-body matrix elements, respectively.
The two solvable $J^\pi=4^+$ and $J^\pi=6^+$ states with seniority $v=4$
are shown with thick lines;
the dashed line corresponds to an intruder level.}
\label{f_ni}
\end{center}
\end{figure}
Let us begin with a discussion of the nickel isotopes
from $^{70}$Ni to $^{76}$Ni;
since virtually nothing is known about the odd-mass nuclei,
let us concentrate on the even-even ones.
The left-hand columns of the spectra in Fig.~\ref{f_ni}
show the observed levels~\cite{Chiara11,Tuli04,Mazzocchi05}.
The nucleus $^{70}$Ni displays, in a single-shell approximation,
a two-neutron-particle spectrum $(\nu1g_{9/2})^2$
with excited states with $J^\pi=2^+$, $4^+$, $6^+$ and $8^+$.
This is indeed found to be the case
except for the additional $J^\pi=(2^+)$ level at 1867~keV.
This level is equally absent in shell-model calculations
in a large basis consisting of the $2p_{1/2}$, $2p_{3/2}$, $1f_{5/2}$ and $1g_{9/2}$ shells
for neutrons and protons~\cite{Lisetskiy04}.
A possible explanation of the 1867~keV level would therefore
seem to require the $1f_{7/2}$ shell
and, in particular, proton excitations across the $Z=28$ shell gap
might lead to low-lying states in $^{70}$Ni.
The nucleus $^{76}$Ni displays a two-neutron-hole spectrum $(\nu1g_{9/2})^{-2}$
with the same yrast sequence as in $^{70}$Ni.
The two-particle and the two-hole spectra
fix the two-body matrix elements $\nu_\lambda$,
or rather the differences $\nu_\lambda-\nu_0$.
This is done at two levels of sophistication
by taking either constant matrix elements
that are the average of those in $^{70}$Ni and $^{76}$Ni (Th1)
or by letting them vary linearly from $^{70}$Ni to $^{76}$Ni (Th2).
In the latter approximation the spectra
of the two-particle and the two-hole nuclei
are exactly reproduced
[except for the intruder $J^\pi=(2^+)$ state in $^{70}$Ni];
the description of the two intermediate isotopes, $^{72}$Ni and $^{74}$Ni,
should be rather accurate, albeit very empirical.

The two solvable $J^\pi=4^+$ and $J^\pi=6^+$ levels with seniority $v=4$
are shown with thick lines in Fig.~\ref{f_ni}.
A noteworthy feature of the calculated spectra of $^{72,74}$Ni
is the occurrence of two levels for $J^\pi=4^+$ and for $J^\pi=6^+$ 
which are very close in energy, especially in the latter case.
This is a direct consequence of the solvability of one member of each doublet
which cannot mix with the close-lying, predominantly $v=2$ state with the same spin.
At least for the $J^\pi=6^+$ levels,
this feature is still clearly present
in the large-scale shell-model calculations of Lisetskiy {\it et al.}~\cite{Lisetskiy04}.

\begin{figure}
\begin{center}
\includegraphics[width=13.8cm]{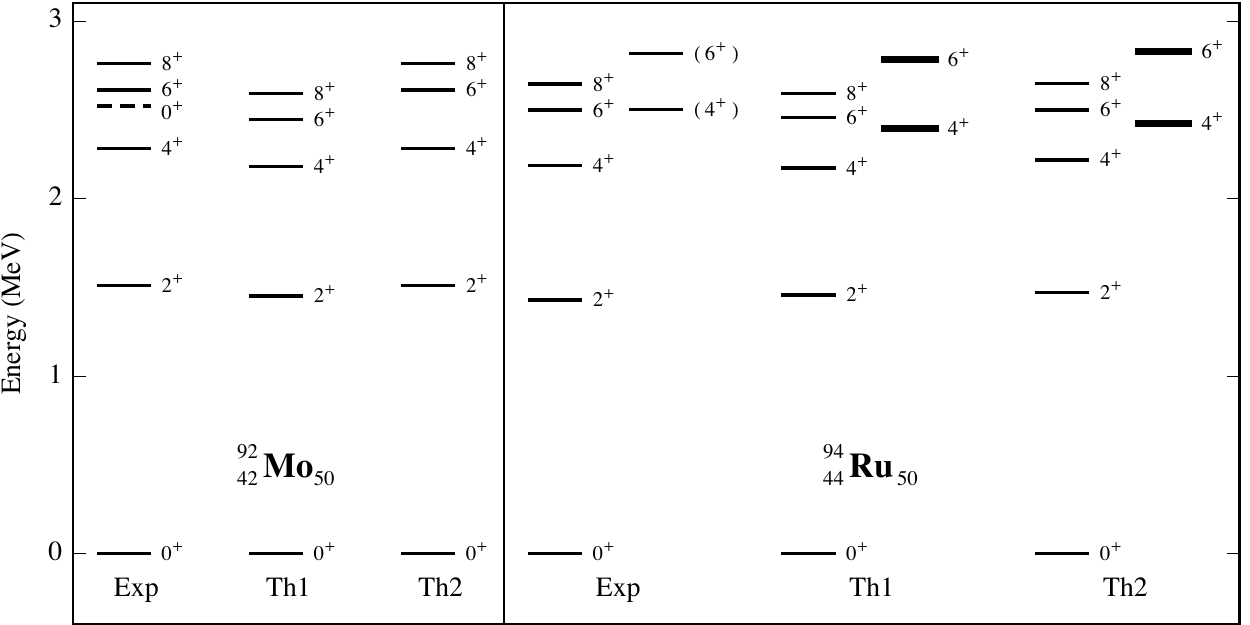}
\includegraphics[width=13.8cm]{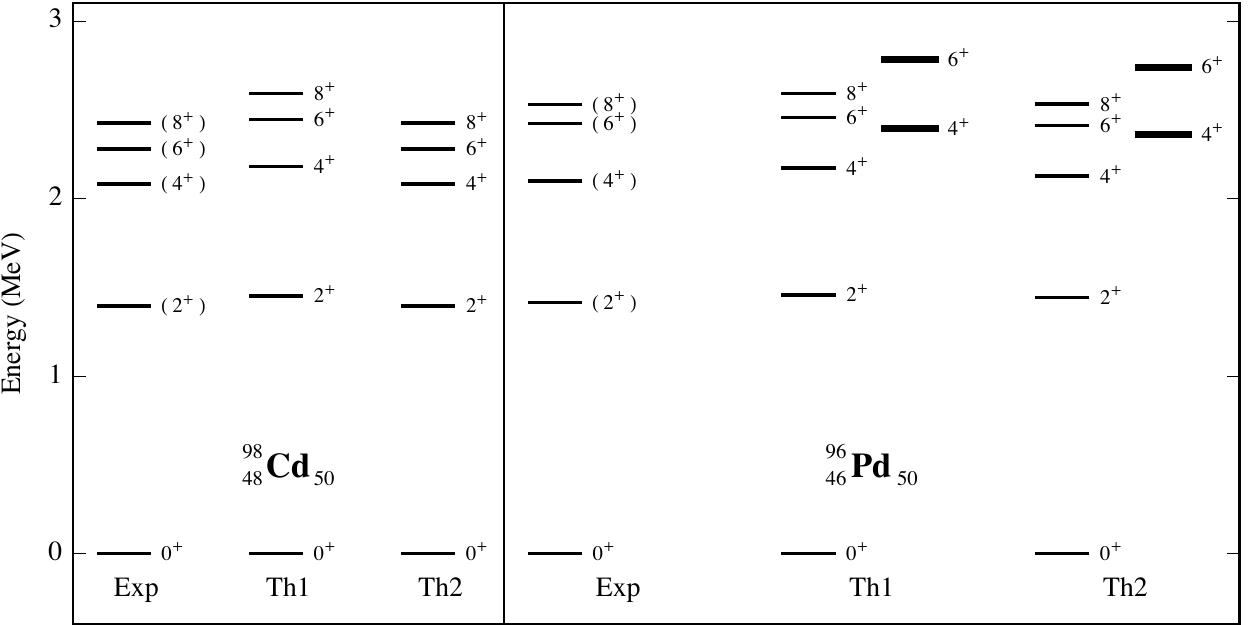}
\caption{The low-energy spectra of the $N=50$ isotones
$^{92}$Mo, $^{94}$Ru, $^{96}$Pd and $^{98}$Cd.
The left-hand columns `Exp' show the observed levels
while the columns `Th1' and `Th2'
contain the results of a $(\pi1g_{9/2})^n$ shell-model calculation
with constant or linearly varying two-body matrix elements, respectively.
The two solvable $J^\pi=4^+$ and $J^\pi=6^+$ states with seniority $v=4$
are shown with thick lines;
the dashed line corresponds to an intruder level.}
\label{f_n50}
\end{center}
\end{figure}
A similar analysis in the same approximation
can be performed for the $N=50$ isotones
with protons in the $\pi1g_{9/2}$ shell.
In Fig.~\ref{f_n50} the results of the calculation
are compared with the observed spectra~\cite{Baglin04,Mills07,Abriola08,Singh03}.
In $^{94}$Ru $(4^+_2)$ and $(6^+_2)$ levels
are observed at energies of 2503 and 2818~keV, respectively~\cite{Mills07};
these are possible candidates for the solvable states
which are calculated at 2422 and 2828~keV in the approximation `Th2'.
It would be of interest to confirm the spin assignment
and to attempt to measure the E2 transition probability between the two levels.

There is a striking difference
between the calculated four-particle (or four-hole) spectra
of the nickel isotopes and those of the $N=50$ isotones:
the solvable $J^\pi=4^+$ and $J^\pi=6^+$ states
are yrast in $^{72}$Ni and $^{74}$Ni
while they are yrare in $^{94}$Ru and $^{96}$Pd.
For this reason one may conjecture
that the observed yrast $(4^+)$ and $(6^+)$ levels in $^{72}$Ni
and the observed yrast $(4^+)$ level in $^{74}$Ni
are the solvable states in question, as is done in Fig.~\ref{f_ni}.

For any reasonable interaction between identical nucleons in a $j=9/2$ shell,
the seniority classification is a good approximation.
This was shown to be true for specific cases by Grawe {\it et al.}~\cite{Grawe02}
but can be argued from general considerations.
Consider as an example the $J^\pi=0^+$ states.
There are two $J^\pi=0^+$ states for four particles in a $j=9/2$ shell
with seniority $v=0$ and $v=4$, respectively.
Introducing the notation
$|0^+_{v=0}\rangle\equiv|(9/2)^4,v=0,J=0\rangle$
and $|0^+_{v=4}\rangle\equiv|(9/2)^4,v=4,J=0\rangle$,
one finds with the help of the results of Sect.~\ref{s_sen4}
the following matrix elements:
\begin{eqnarray}
\langle0^+_{v=0}|\hat V|0^+_{v=0}\rangle&=&
\frac{8}{5}\nu_0+\frac{1}{2}\nu_2+\frac{9}{10}\nu_4+\frac{13}{10}\nu_6+\frac{17}{10}\nu_8,
\nonumber\\
\langle0^+_{v=0}|\hat V|0^+_{v=4}\rangle&=&
\frac{-65\nu_2+315\nu_4-403\nu_6+153\nu_8}{10\sqrt{429}},
\nonumber\\
\langle0^+_{v=4}|\hat V|0^+_{v=4}\rangle&=&
\frac{13}{66}\nu_2+\frac{735}{286}\nu_4+\frac{961}{330}\nu_6+\frac{459}{1430}\nu_8.
\label{e_j9mat0}
\end{eqnarray}
The extent of the breaking of seniority
depends on the size of the off-diagonal matrix element
divided by the difference between the diagonal ones.
For any reasonable choice of interaction this ratio is small.
For example, for any of the interactions fitted
to the nickel isotopes or the $N=50$ isotones,
the differences between the exact and the diagonal energies
are less than 2 keV
and the admixtures of seniority $v=4$ in the ground state
do not exceed 0.1\% in amplitude.
A similar argument applies to the $J^\pi=2^+$ and $J^\pi=8^+$ states.

The proof that seniority mixing is negligible
for all $J^\pi=4^+$ and $J^\pi=6^+$ states
of a four-particle $j=9/2$ system is more subtle.
There are {\em three} states for each of these angular momenta,
two of which, with seniority $v=2$ and $v=4$, are close in energy
and could possibly strongly mix.
However, the seniority $v=4$ members of the closely-spaced doublets
are precisely the solvable $J^\pi=4^+$ and $J^\pi=6^+$ states
discussed in Sect.~\ref{s_sen4},
and they conserve seniority for any interaction.
As a consequence, breaking of seniority
only arises through mixing of the seniority $v=2$
and the higher-lying seniority $v=4$ state
and, by the same argument as above,
this mixing is found to be small.

The conclusion of this discussion is that,
for any reasonable two-body interaction,
seniority is a good quantum number
for all states in a $j=9/2$ shell.
Exact energies obtained from a diagonalization
are close to the approximate seniority formulas,
which for the $J^\pi=4^+$ and $J^\pi=6^+$ states with seniority $v=2$
are given by
\begin{eqnarray}
E[(9/2)^4,v=2,J=4]&=&
\frac{3}{5}\nu_0+\frac{67}{99}\nu_2+\frac{746}{715}\nu_4+\frac{1186}{495}\nu_6+\frac{918}{715}\nu_8,
\nonumber\\
E[(9/2)^4,v=2,J=6]&=&
\frac{3}{5}\nu_0+\frac{34}{99}\nu_2+\frac{1186}{715}\nu_4+\frac{658}{495}\nu_6+\frac{1479}{715}\nu_8.
\label{e_j9e46}
\end{eqnarray}
Comparison of these expressions with the corresponding ones
for the solvable $J^\pi=4^+$ and $J^\pi=6^+$ states with seniority $v=4$, Eq.~(\ref{e_j9energ4}),
makes it clear that the lowering of the solvable states in $^{72,74}$Ni
is associated with the low excitation energy of the $J^\pi=2^+$ level
in the two-particle and two-hole nuclei $^{70}$Ni and $^{76}$Ni.
In the corresponding $N=50$ isotones, $^{92}$Mo and $^{98}$Cd,
the $J^\pi=2^+$ level is at higher energy
and, because the coefficient of $\nu_2$ in Eq.~(\ref{e_j9energ4})
is larger than the one in Eq.~(\ref{e_j9e46}),
this results in a higher excitation energy of both solvable states
in $^{94}$Ru and $^{96}$Pd.

\begin{figure}
\begin{center}
\includegraphics[width=8cm]{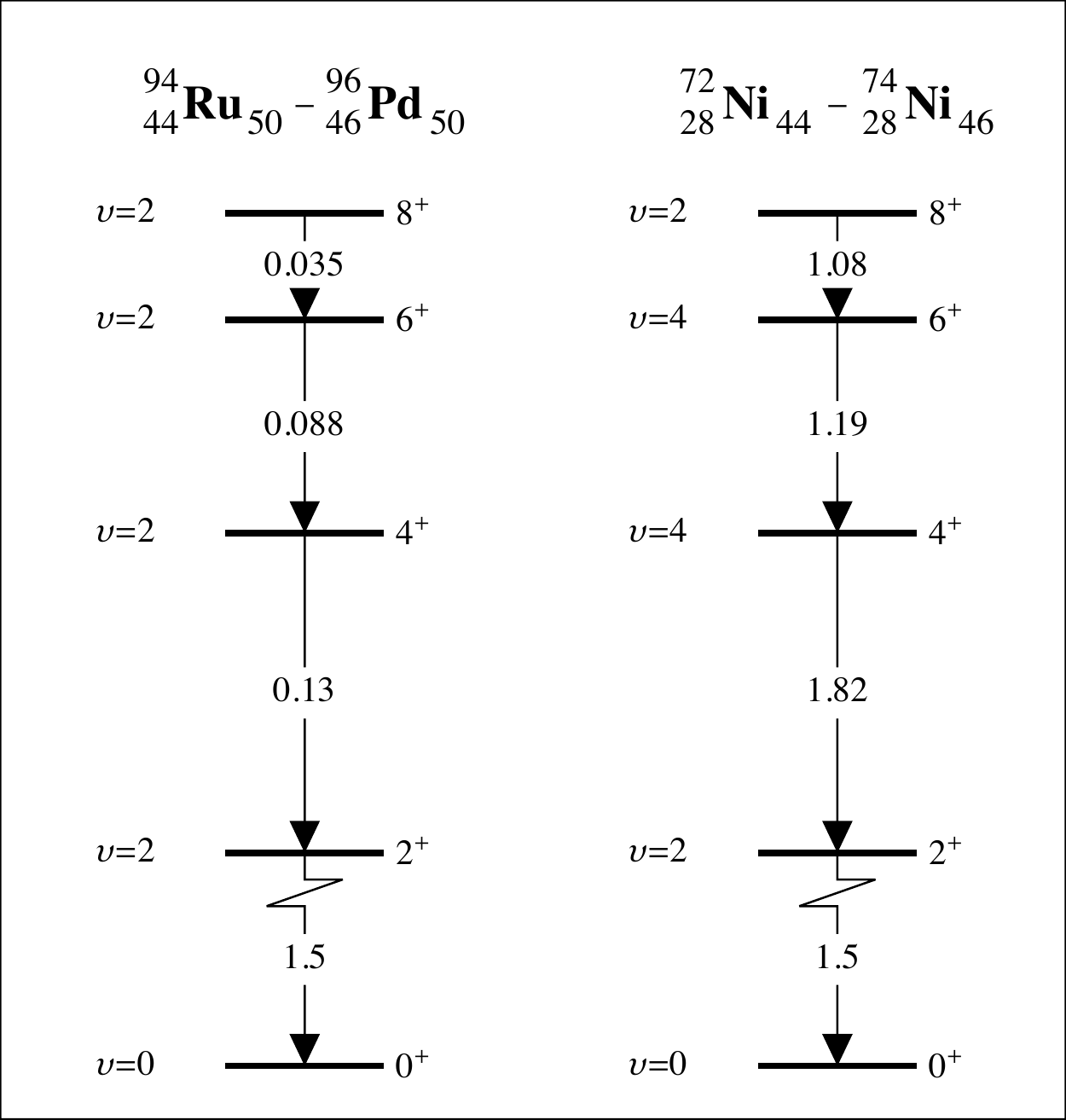}
\caption{
E2 decay in the $(9/2)^4$ system
as expected in the $N=50$ isotones (left)
and the $Z=28$ nickel isotopes (right).
The numbers between the levels denote $B$(E2) values
expressed in units of $B({\rm E2};2^+_1\rightarrow0^+_1)$
of the two-particle system
and obtained with a seniority-conserving interaction.}
\label{f_be2}
\end{center}
\end{figure}
Partial seniority conservation sheds also
some new light on the existence of seniority isomers~\cite{Isacker08,Isacker11}.
Figure~\ref{f_be2} illustrates
the E2 decay for four particles in the $j=9/2$ shell
as obtained with a seniority-conserving interaction.
On the left-hand side it shows the `typical' decay
with very small $B$(E2) values between states with seniority $v=2$
which is characteristic of the seniority classification
in nuclei near mid-shell ($N\approx j+1/2$)
and which is at the basis of the explanation of seniority isomers~\cite{Grawe97}.
This situation applies to $^{94}$Ru and $^{96}$Pd
where the states with seniority $v=2$ are yrast.
Another condition, necessary for the existence of the $J^\pi=8^+$ state as an isomer,
is that it should occur {\em below} the solvable $J^\pi=6^+$,
which can be easily verified by comparing its energy, Eq.~(\ref{e_j9energ4}),
with the one of the $J^\pi=8^+$ state,
in a seniority approximation given by
\begin{equation}
E[(9/2)^4,v=2,J=8]=
\frac{3}{5}\nu_0+\frac{6}{11}\nu_2+\frac{486}{715}\nu_4+\frac{87}{55}\nu_6+\frac{1854}{715}\nu_8.
\label{e_j9e8}
\end{equation}
On the right-hand side of Fig.~\ref{f_be2} is shown
the E2 decay pattern as it is calculated in $^{72,74}$Ni.
As argued above, it can be expected in these isotopes
that the yrast $J^\pi=4^+$ and $J^\pi=6^+$ levels have seniority $v=4$,
and this drastically alters the E2 decay pattern in the yrast band.
As a consequence,
unless the $J^\pi=8^+$ and $J^\pi=6^+$ levels are very close in energy,
the former is unlikely to be isomeric.

\begin{figure}
\begin{center}
\includegraphics[width=10cm]{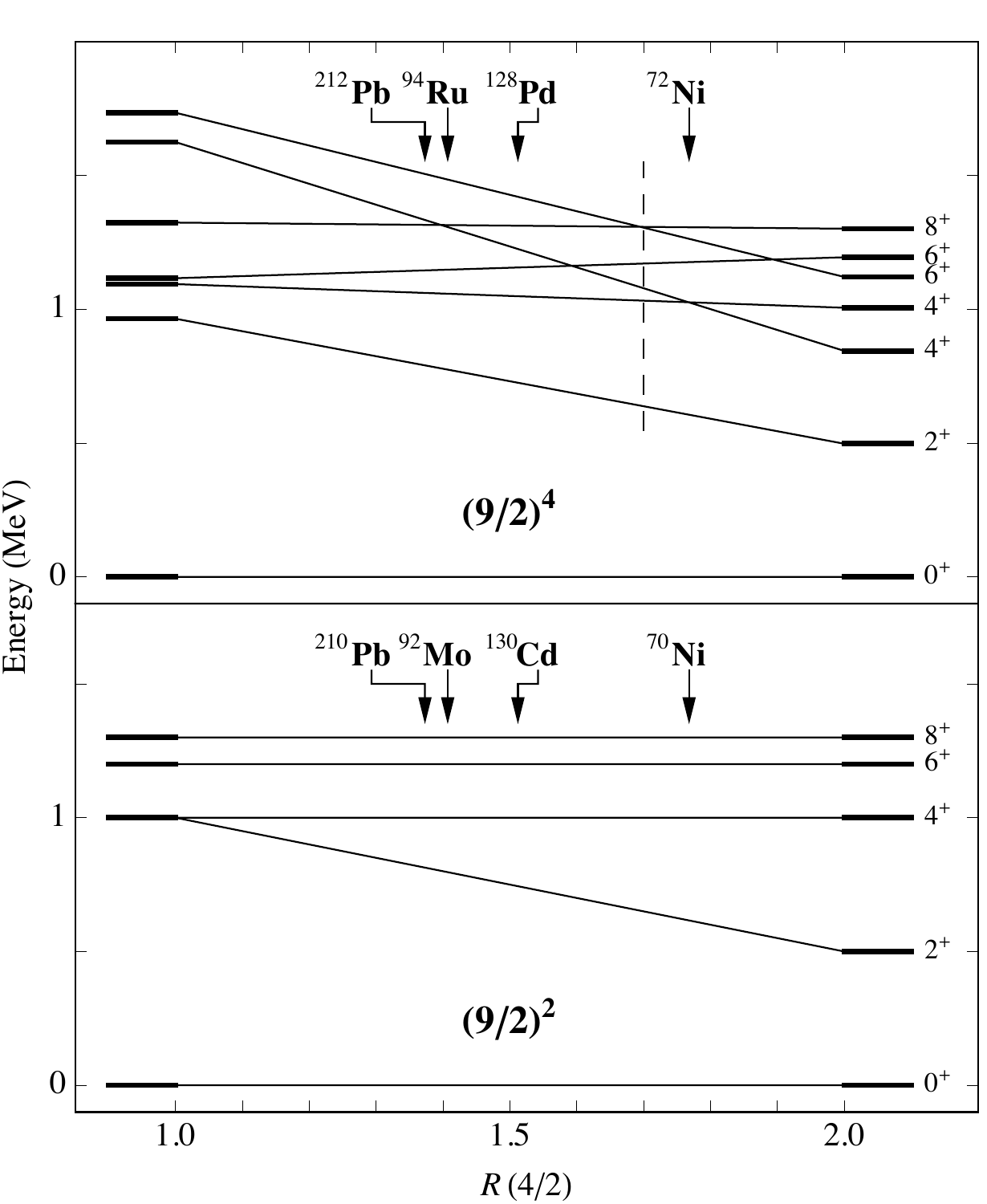}
\caption{
Schematic illustration of the effect of $R(4/2)$,
the ratio of excitation energies of the yrast $J^\pi=4^+$ and $J^\pi=2^+$ levels
in the two-particle spectrum (bottom),
on the properties of the four-particle spectrum (top).
The arrows indicate the positions
of a number of isotopes according to their $R(4/2)$ ratio
in the two-particle (or two-hole) spectra,
and the corresponding four-particle (or four-hole) isotopes.}
\label{f_r42}
\end{center}
\end{figure}
The results of this analysis are schematically summarized in Fig.~\ref{f_r42}
which shows the evolution of the two- and four-particle spectra
as a function of $R(4/2)$,
the ratio of excitation energies of the yrast $J^\pi=4^+$ and $J^\pi=2^+$ levels
in the two-particle spectrum.
This ratio is assumed in the figure to vary between two extreme values,
$R(4/2)=1$ and $R(4/2)=2$,
beyond which a seniority classification certainly is not any longer a reasonable approximation.
The excitation energies of the yrast levels in the two-particle spectrum
with $J^\pi=4^+$, $J^\pi=6^+$ and $J^\pi=8^+$
are assumed to be constant.
In the upper panel of Fig.~\ref{f_r42} is shown
the evolution of levels of the four-particle spectrum
as a function of $R(4/2)$ in the two-particle spectrum.
If effects of seniority mixing (which are small) are neglected,
excitation energies vary linearly with $R(4/2)$.
Not surprisingly, the excitation energy
of the $J^\pi=2^+$ level in the four-particle spectrum
approximately drops by a factor two as $R(4/2)$ changes from 1 to 2
but the other levels with seniority $v=2$ remain approximately constant in energy.
The biggest change, however, occurs for the two solvable levels with seniority $v=4$
to the extent that they cross some of the levels with seniority $v=2$.
The dashed line indicates the point where the $J^\pi=6^+$ level with seniority $v=4$
crosses the $J^\pi=8^+$ level with seniority $v=2$;
the latter level is expected to be isomeric
for smaller values of $R(4/2)$.
The arrows indicate the positions on this diagram
of a number of isotopes according to their $R(4/2)$ ratio
in the two-particle (or two-hole) spectra.
Extrapolation to the four-particle (or four-hole) spectra
then leads to the conclusion that no $J^\pi=8^+$ isomer
should exist in $^{72}$Ni (nor in $^{74}$Ni)
while they should occur in $^{212}$Pb (and $^{214}$Pb),
$^{94}$Ru (and $^{96}$Pd),
as well as in $^{128}$Pd (and $^{126}$Ru).

In a less schematic analysis
the exact positions of the $J^\pi=6^+$ and $J^\pi=8^+$ levels
of the two-particle spectrum should be taken
instead of the plausible but somewhat arbitrary values of Fig.~\ref{f_r42}.
The influence of these energies in this analysis is weak, however,
and the same conclusion is obtained with the correct energies,
that is, the $J^\pi=8^+$ levels of the four-particle (or four-hole) nuclei
are isomeric except those in $^{72}$Ni and $^{74}$Ni.

Besides the nickel isotopes and the $N=50$ isotones,
already discussed in the preceding paragraphs,
Fig.~\ref{f_r42} predicts properties of the neutron-rich lead isotopes ($Z=82$)
and the proton-poor $N=82$ isotones.
The two-particle nucleus $^{210}$Pb is particularly stiff,
characterized by a low ratio $R(4/2)=1.37$
and therefore $J^\pi=8^+$ seniority isomers
should exist in $^{212}$Pb and $^{214}$Pb.
This is indeed confirmed by recent experiments~\cite{Gottardo12}.

\begin{figure}
\begin{center}
\includegraphics[width=10cm]{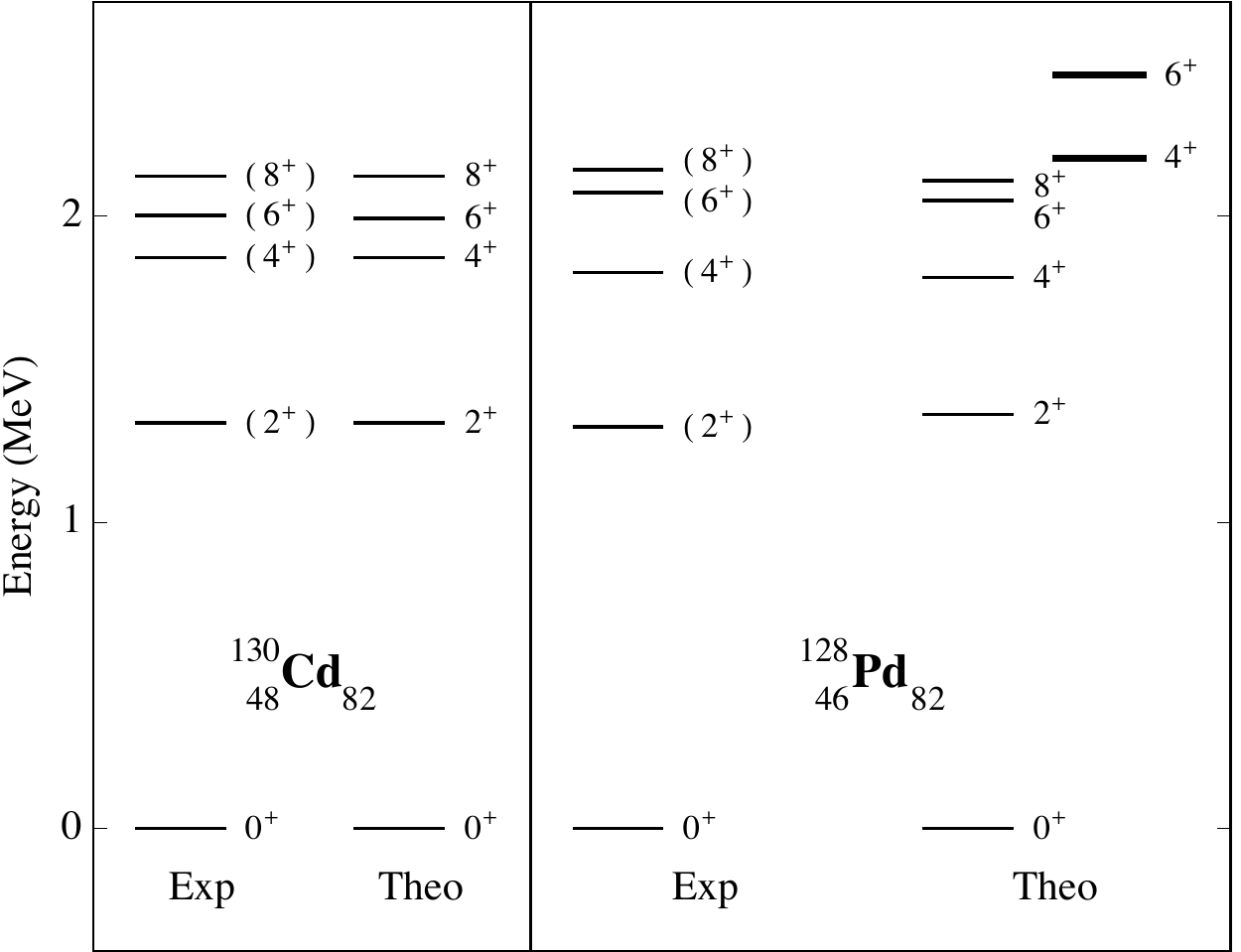}
\caption{The low-energy spectra of the isotopes $^{130}$Cd and $^{128}$Pd.
The left-hand columns `Exp' show the observed levels
while the columns `Theo' contain the results of a $(\pi1g_{9/2})^n$ shell-model calculation
with constant two-body matrix elements.
The two solvable $J^\pi=4^+$ and $J^\pi=6^+$ states with seniority $v=4$
are shown with thick lines.}
\label{f_cd130pd128}
\end{center}
\end{figure}
Another interesting application concerns
the existence of a $J^\pi=8^+$ seniority isomer in $^{128}$Pd.
An isomer in this extremely neutron-rich palladium isotope ($Z=46$)
was detected at the RIBF facility at RIKEN
with a delayed-coincidence technique
between the identified radioactive ion
and gamma rays de-exciting the isomeric state
after passing through the spectrometer~\cite{Watanabe13}.
The two-hole spectrum of $^{130}$Cd is well known
and contains a $J^\pi=8^+$ isomer with a half-life of 0.22(3)~$\mu$s~\cite{Jungclaus07}.
On the basis of these properties of $^{130}$Cd,
a simple prediction can be made of those of $^{128}$Pd (see Fig.~\ref{f_cd130pd128})
with a $J^\pi=8^+$ level at 2116~keV
which turns out to be isomeric
since it occurs below the $J^\pi=6^+$ state with seniority $v=4$ at 2460~keV
and just above the $J^\pi=6^+$ state with seniority $v=2$ at 2051~keV.
The half-life $T_{1/2}({\rm ^{128}Pd};8^+_1)$ of the $J^\pi=8^+$ level in $^{128}$Pd
can be estimated from
\begin{equation}
\frac{T_{1/2}({\rm ^{128}Pd};8^+_1)}{T_{1/2}({\rm ^{130}Cd};8^+_1)}=
\frac{1+\alpha_{\rm Cd}(128)}{1+\alpha_{\rm Pd}(75)}
\left(\frac{128}{75}\right)^5
\frac{B({\rm E2};8^+_1\rightarrow6^+_1)_{\rm ^{130}Cd}}{B({\rm E2};8^+_1\rightarrow6^+_1)_{\rm ^{128}Pd}},
\label{e_t1/2a}
\end{equation}
where $\alpha_{\rm X}(E_\gamma)$ is the internal electron conversion coefficient
for a transition with energy $E_\gamma$ (in keV) in the element X.
The gamma-ray energies for the $8^+_1\rightarrow6^+_1$ transitions
in $^{128}$Pd and $^{130}$Cd, $E_\gamma=75$ and 128,
are taken from experiment~\cite{Watanabe13,Jungclaus07}.
In the seniority scheme the following relations are valid (see Fig.~\ref{f_be2}):
\begin{eqnarray}
B({\rm E2};8^+_1\rightarrow6^+_1)_{\rm ^{130}Cd}&=&
0.318\,B({\rm E2};2^+_1\rightarrow0^+_1)_{\rm ^{130}Cd},
\nonumber\\
B({\rm E2};8^+_1\rightarrow6^+_1)_{\rm ^{128}Pd}&=&
0.035\,B({\rm E2};2^+_1\rightarrow0^+_1)_{\rm ^{130}Cd},
\label{e_be2a}
\end{eqnarray}
which, together with the values for the conversion coefficients
as obtained from the database BrIcc~\cite{Kibedi08},
$\alpha_{\rm Cd}(128)=0.621$ and $\alpha_{\rm Pd}(75)=3.90$,
leads to the estimate $T_{1/2}({\rm ^{130}Cd};8^+_1)\approx9.6$~$\mu$s,
which is reasonably close to the observed value of 5.8(8)~$\mu$s~\cite{Watanabe13}.
A possible source of error in the theoretical estimate
is the smallness of the $B({\rm E2};8^+_1\rightarrow6^+_1)$ value in $^{128}$Pd,
which is sensitive to small admixtures in the wave functions.

One should be aware of the limits of validity
of the simple estimates given on the basis of seniority.
To illustrate this point,
let us return to the example of the nickel isotopes.
Besides the half-life of the $J^\pi=8^+$ isomer in $^{70}$Ni,
also the Coulomb-excitation probability of the first-excited $J^\pi=2^+$ state
is known in this nucleus,
leading to a $B({\rm E2};2^+_1\rightarrow0^+_1)$ value
of $172(28)$~$e^2$fm$^4$~\cite{Perru06}.
On the basis of the seniority relation between the $B$(E2) values
for the $2^+_1\rightarrow0^+_1$ and $8^+_1\rightarrow6^+_1$ transitions,
see the first equation of Eq.~(\ref{e_be2a}),
a half-life $T_{1/2}({\rm ^{70}Ni};8^+_1)=73(12)$~ns is deduced,
a factor three shorter than what is observed.
In other words, the $2^+_1\rightarrow0^+_1$ E2 transition
is faster by a factor three than expected on the basis of seniority,
indicating that the $J^\pi=2^+_1$ state has a collective structure
that goes beyond a pure $\nu1g_{9/2}$ shell.

One can push this argument further
and estimate the half-life of the $J^\pi=(8^+)$ level in $^{72}$Ni.
Because of the structure of the solvable $J^\pi=6^+$ state with seniority $v=4$,
see the second equation of Eq.~(\ref{e_j9wave}),
the $8^+_1\rightarrow6^+_1$ E2 transition
arguably can be expected to be collectively enhanced as well.
It should therefore be estimated
from the $2^+_1\rightarrow0^+_1$
and {\em not} from the $8^+_1\rightarrow6^+_1$ E2 transition in $^{70}$Ni,
leading to
\begin{equation}
\frac{T_{1/2}({\rm ^{72}Ni};8^+_1)}{T_{1/2}({\rm ^{70}Ni};2^+_1)}=
\frac{1+\alpha_{\rm Ni}(1260)}{1+\alpha_{\rm Ni}(199)}
\left(\frac{1260}{199}\right)^5
\frac{B({\rm E2};2^+_1\rightarrow0^+_1)_{\rm ^{70}Ni}}{B({\rm E2};8^+_1\rightarrow6^+_1)_{\rm ^{72}Ni}}.
\label{e_t1/2b}
\end{equation}
From the relation (see Fig.~\ref{f_be2})
\begin{equation}
B({\rm E2};8^+_1\rightarrow6^+_1)_{\rm ^{72}Ni}=
1.08\,B({\rm E2};2^+_1\rightarrow0^+_1)_{\rm ^{70}Ni},
\label{e_be2b}
\end{equation}
and the half-life $T_{1/2}({\rm ^{70}Ni};2^+_1)=1.6(3)$~ps,
deduced from the $B({\rm E2};2^+_1\rightarrow0^+_1)$ value,
one obtains the estimate $T_{1/2}({\rm ^{72}Ni};8^+_1)\approx14$~ns,
which is consistent with the current upper limit of 20~ns~\cite{Chiara11}.

\subsection{Seniority and one-nucleon transfer}
\label{ss_transfer}
The energy spectrum of four identical particles (or holes)
in a $j=9/2$ shell will, for any reasonable nuclear interaction,
display two $J^\pi=4^+$ and two $J^\pi=6^+$ levels which are close in energy.
Despite this closeness in energy all states retain their character.
In fact, under the assumption of a pure $(9/2)^4$ configuration,
the partial conservation of seniority
leads to one level with exact seniority $v=4$
and the other with approximate seniority $v\approx2$
as it mixes (but only weakly) with the other state with seniority $v=4$ at higher energy.

\begin{figure}
\includegraphics[width=6.9cm]{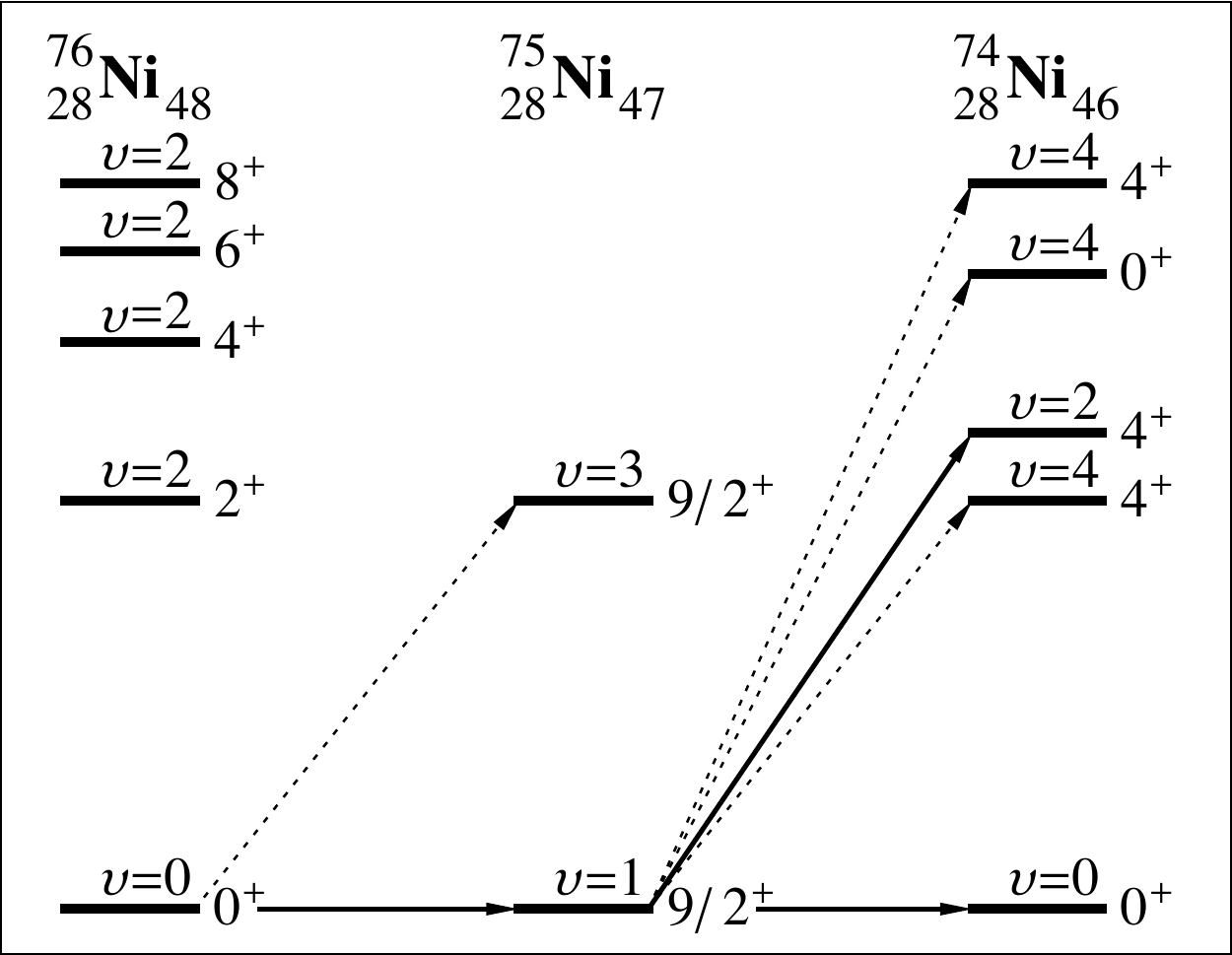}
\includegraphics[width=6.9cm]{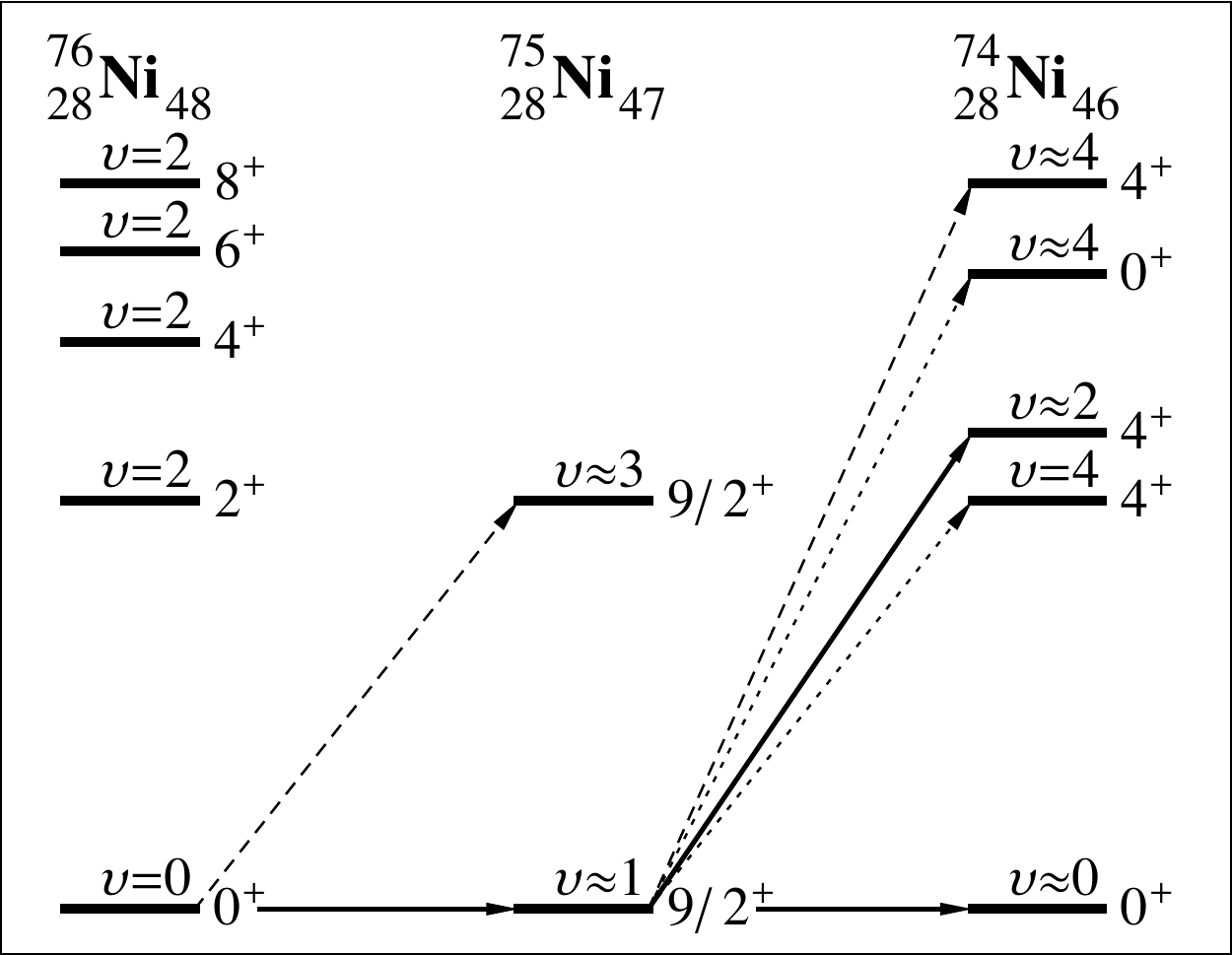}
\caption{
Selection rules for one-particle transfer in a $j=9/2$ shell.
In the left panel seniority is an exact quantum number for all states
while in the right panel it is broken for most states,
except for the two-particle states
and for the four-particle $J^\pi=4^+$ state with seniority $v=4$.
Full lines indicate transitions allowed in one-particle transfer,
dotted lines correspond to forbidden transitions
and dashed lines are transitions which arise due to seniority mixing
and therefore are weak but not exactly zero.
Example nuclei corresponding to the two-, three- and four-particle systems
are indicated on top.
Energies of levels are not drawn to scale.}
\label{f_trani}
\end{figure}
Partial conservation of seniority should therefore
have consequences with regard to one-particle transfer.
Since the seniority of a single particle is $v=1$,
the selection rule associated with this reaction is $\Delta v=\pm1$.
If seniority is conserved for {\em all} eigenstates in a $j=9/2$ shell,
the intensities of the transfer
from a two- to a three-particle system
and from a three- to a four-particle system
are therefore as indicated in the left panel of Fig.~\ref{f_trani}:
they vanish exactly if $\Delta v\neq\pm1$.
For clarity only the selection rules for the $J^\pi=4^+$ states
are indicated in Fig.~\ref{f_trani}
but the same ones are valid for the $J^\pi=6^+$ states.

If, more realistically, a general and arbitrary interaction
is taken in the $j=9/2$ shell,
the intensity pattern shown in the left panel of Fig.~\ref{f_trani} applies.
As expected, since most of the levels
do not carry any longer exact seniority,
some of the forbidden transitions become allowed;
their intensities remain small, however,
since the seniority mixing is expected to be weak.
More surprisingly, two of the one-particle transfer intensities
remain {\em exactly} zero in spite of the seniority mixing.
The explanation of these selection rules is intricate
and related to the geometry of the $j=9/2$ shell.

Firstly, the one-particle transfer
from the $J^\pi=9/2^+$ three-particle ground state with seniority $v\approx3$
to the second $J^\pi=0^+$ four-particle state with seniority $v\approx4$
remains exactly forbidden.
The seniority mixing for the $J^\pi=9/2^+$ states
is derived from the matrix
\begin{eqnarray}
\langle9/2^+_{v=1}|\hat V|9/2^+_{v=1}\rangle&=&
\frac{4}{5}\nu_0+\frac{1}{4}\nu_2+\frac{9}{20}\nu_4+\frac{13}{20}\nu_6+\frac{17}{20}\nu_8,
\nonumber\\
\langle9/2^+_{v=1}|\hat V|9/2^+_{v=3}\rangle&=&
\frac{-65\nu_2+315\nu_4-403\nu_6+153\nu_8}{20\sqrt{429}},
\nonumber\\
\langle9/2^+_{v=3}|\hat V|9/2^+_{v=3}\rangle&=&
\frac{13}{132}\nu_2+\frac{735}{572}\nu_4+\frac{961}{660}\nu_6+\frac{459}{2860}\nu_8.
\label{e_j9mat9}
\end{eqnarray}
It is seen that this mixing matrix is proportional
to the one for the $J^\pi=0^+$ states given in Eq.~(\ref{e_j9mat0}).
The mixings cancel as a result
and the seniority selection rule remains exact.

Secondly, the one-particle transfer
from the $J^\pi=9/2^+$ ground state of the three-particle system
to the solvable $J^\pi=4^+$ and $J^\pi=6^+$ four-particle state with seniority $v=4$
also remains exactly forbidden,
despite $v=3$ admixtures in the former state.
The reason in this case is that
\begin{equation}
[j^3(v=3,J_3=j)jJ|\}j^4,v=4,{\rm s},J]=0,
\end{equation}
for $j=9/2$ and for the solvable four-particle states
with $J=4$ and 6~\cite{Zamick08,Qi11}.

In summary, the interesting aspect of these results
is that two $J^\pi=4^+$ states and two $J^\pi=6^+$ states
are predicted to be close in energy
but that they should be differently excited in a one-particle transfer reaction.
To test whether these schematic predictions
remain valid in a more realistic scenario,
they should be compared with the results
of a large-scale shell-model (LSSM) calculation.
This has been done for the neutron-rich nickel isotopes
with the conclusion that these characteristic features
are still present in the LSSM calculation
for the $J^\pi=6^+$ but not for the $J^\pi=4^+$ states~\cite{Isacker12}.

\section{Conclusions}
\label{s_conc}
In this paper a review
of the seniority quantum number in many-body systems was given.
The analysis was carried out for bosons and fermions simultaneously
but was restricted to identical particles occupying a single shell
(or, alternatively, particles with the same spin).
The conditions of complete solvability were shown to be more restrictive
than those for the conservation of seniority.
The {\em partial} conservation of seniority
was shown to be a peculiar property of spin-9/2 fermions
but prevalent in systems of interacting bosons of any spin.
Partial conservation of seniority was shown
to be at the basis of the existence of seniority isomers
which are frequently found in semi-magic nuclei,
and to give rise to selection rules in one-nucleon transfer reactions
that have yet to be tested experimentally.

A general result of the present analysis
is the proof that any system of interacting bosons with spin $\ell\leq2$ is integrable
and that its spectrum in energy is analytically available
for any number of bosons.
This property has been exploited in the discussion
of Bose-Einstein condensates
consisting of atoms with hyperfine spins $f=1$~\cite{Law98}
and $f=2$~\cite{Isacker07,Uchino08,He11}.
A more subtle property is the partial conservation of seniority
in systems of interacting bosons with spin $\ell>2$.
The consequences of this finding for Bose-Einstein condensates
are still to be explored.

This work, elementary though it may be,
paves the way for further studies which can be carried out along similar lines.
Possible generalizations concern systems of
(i) non-identical particles
and (ii) particles distributed over several shells.
Open-shell atomic nuclei provide examples of fermionic systems of type (i).
The isoscalar component of the nuclear interaction, however, strongly breaks seniority
and, therefore, the study of seniority in such systems has seemed irrelevant so far.
Given the current interest in two-component Bose-Einstein condensates,
the problem should be revisited for bosons.
By considering semi-magic nuclei as examples of systems of type (ii),
they can be treated more realistically
than with the approach advocated in the present paper.
Moreover, interesting formal questions can be explored
concerning connections with generalized seniority~\cite{Talmi71}
as well as with integrable Richardson-Gaudin models~\cite{Dukelsky04}.
These problems are currently under study
and will be the material of the subsequent papers in this series.

\section*{Acknowledgements}
Part of this work was done in collaboration with
Igor \v{C}elikovi\'c and Larry Zamick.
We wish to thank
Alex Brown,
Ami Leviatan,
Igal Talmi
and John Wood
for illuminating discussions at various stages of this work.

\end{document}